\documentclass[a4paper,11pt]{article}
\usepackage{jheppub}
\usepackage{amsmath,amssymb,amsfonts,amsthm,amscd,graphicx,physics}
\usepackage{xcolor}
\usepackage{hyperref}
\usepackage{tikz}
    \usepackage{tkz-euclide}
    \usetikzlibrary{arrows,calc,patterns}
\usepackage[utf8]{inputenc}
\usepackage{geometry}
\usepackage{mathrsfs}
\usepackage{dsdshorthand}
\usepackage{changepage}
\usepackage[normalem]{ulem}

\usepackage{setspace}
\usepackage{CJKutf8}
\usepackage{subcaption}

\title{\begin{center}
        Consistent Evaluation of the No-Boundary Proposal
\end{center}}  

\author[1]{Ahmed I. Abdalla,}
\author[1]{Stefano Antonini,}
\author[1]{Raphael Bousso,}
\author[1]{Luca V.~Iliesiu,}
\author[2]{Adam Levine,}
\author[3]{Arvin Shahbazi-Moghaddam}

\affiliation[1]{Leinweber Institute for Theoretical Physics and Department of Physics, University of California, Berkeley, California 94720, U.S.A.}

\affiliation[2]{Center for Theoretical Physics -- a Leinweber Institute, Massachusetts Institute of Technology, \\Cambridge, MA 02139, USA}

\affiliation[3]{Leinweber Institute for Theoretical Physics, Stanford University, Stanford, CA 94305}

\emailAdd{aiabdalla@berkeley.edu}
\emailAdd{santonini@berkeley.edu}
\emailAdd{bousso@berkeley.edu}
\emailAdd{liliesiu@berkeley.edu}
\emailAdd{arlevine@mit.edu}
\emailAdd{arvinshm@gmail.com}

\abstract{
We revisit the Hartle-Hawking no-boundary proposal. To extract probabilities, one must use the gravitational path integral (GPI) to compute not only the no-boundary amplitude, but also the norms by which its square is divided. We find that this dramatically alters predictions: the probability for any closed universe
is either nearly 1, or exactly 1. That is, in the Hilbert space of closed universes defined by the GPI, the states of interest in cosmology are all nearly parallel to the Hartle-Hawking state up to nonperturbative corrections in $G_N^{-1}$. We also consider a statistical interpretation of the GPI, as an average of arbitrary products of amplitudes. We find that all amplitudes are exactly 1 in this case, consistent with recent arguments that the statistical approach to the GPI with a closed boundary computes an average over one-dimensional Hilbert spaces. As an example, we illustrate the consistent evaluation of the no-boundary proposal in inflationary cosmology.}

\date{\today}

\def\HH{\ket{\varnothing}}

\newcommand{\inlinefig}[2][8]{
    \raisebox{-0.37\totalheight}{\includegraphics[height=#1\fontcharht\font`\B]{#2}}
}

\begin{document}

\maketitle

\parskip=3pt

\section{Introduction}

The gravitational path integral (GPI) is the map 
\begin{equation}\label{eq:definition-G}
G(J) \equiv \int_J \cD g_{\mu\nu} \cD \phi\, e^{-I(g_{\mu\nu},\phi)}~,
\end{equation}
from boundary conditions $J$ specified on a closed manifold $\Sigma$ of dimension $d$, to the complex numbers. $\Sigma$ need not be connected. The boundary conditions may consist of a nondegenerate induced metric and the field values on $\Sigma$ (Dirichlet boundary conditions \cite{Gibbons:1976ue,York:1972sj}),
\be\label{eq:jh}
J = (h_{ij}|_{\Sigma}, \phi|_{\Sigma},\ldots),
\ee
or of other well-posed choices, such as fixing the Brown-York stress-tensor 
(Neumann boundary conditions \cite{Krishnan:2016mcj}) or the trace of the extrinsic curvature and the conformal class of the induced metric (conformal boundary conditions \cite{Galante:2025emz,Banihashemi:2025qqi}). The integral \eqref{eq:definition-G} is taken over all physically inequivalent $d+1$ dimensional complex manifolds $M$ with metrics $g_{\mu\nu}$ and matter field configurations $\phi$ that match the boundary values $J$ on $\Sigma = \partial M$. The topology of $M$ is arbitrary; in particular, $M$ need not be connected, and $M$ may contain connected components that have no boundary at all.
$I(g_{\mu\nu},\phi)$ denotes the Euclidean action for gravity and the matter fields.\footnote{The full path integral is under good computational control only in low-dimensional toy models. More generally, we rely on a saddlepoint approximation that can be employed whenever higher loop corrections are suppressed in $L^{d-2}/G_N$, where $L \gg G_N^{1/(d-2)}$ is the characteristic length-scale for each saddle point. The saddlepoint approximation, in turn, relies on a contour prescription that determines which saddles contribute to $G(J)$.} 

The boundary conditions $J$ specify half of the phase-space data on a closed universe $\Sigma$. Thus $J$ defines a quantum state $\ket J$. The central assumption underlying the path integral approach to quantum cosmology \cite{Hawking:1980gf,Hawking:1981gb,hartle_wave_1983} 
is that \emph{the GPI with two closed boundaries defines an unnormalized inner product}: 
\begin{equation}\label{eq:uip}
    \braket{J_1}{J_2}= G(J_1^*,J_2)\equiv G(J_1^*\cup J_2)~.
\end{equation} 
Above, the star indicates an involution of the boundary conditions so that \eqref{eq:uip} is anti-linear in $J_1$ and linear in $J_2$.\footnote{See \cite{witten_bras_2025} and footnote 3 of \cite{harlow_gauging_2025} for additional details.} The normalized inner product is therefore given by
\begin{equation}\label{eq:ip}
    \frac{\braket{J_1}{J_2}}{\braket{J_1}{J_1}^{1/2}\braket{J_2}{J_2}^{1/2}} = \frac{G(J_1^*,J_2)}{G(J_1^*,J_1)^{1/2}G(J_2^*,J_2)^{1/2}}\,.
\end{equation}
Below, we will also discuss a different ``statistical'' or ``ensemble'' interpretation of the GPI \cite{Saad:2018bqo,Penington:2019kki,saad_jt_2019,Bousso:2019ykv,marolf_transcending_2020,Sasieta:2022ksu,Balasubramanian:2022gmo,Chandra:2022bqq,DiUbaldo:2023qli,Jafferis:2025jle,Chandra:2023dgq,marolf_nature_2024,usatyuk_closed_2024,Usatyuk:2024isz,abdalla_gravitational_2025,harlow_quantum_2025}.  Recent papers on the GPI approach to quantum cosmology include \cite{Maldacena_2004,McInnes_2004,Cooper_2019,Antonini:2019qkt,McNamara:2020uza, Chen:2020tes,VanRaamsdonk:2020tlr,VanRaamsdonk:2021qgv,Antonini:2022blk,Antonini:2022ptt,sahu2023bubblescosmologyadscft,Betzios:2024oli,sahu2024holographicblackholecosmologies,Antonini:2023hdh,Antonini:2024mci,Engelhardt:2025vsp,Antonini:2025ioh,Liu:2025cml,Gesteau:2025obm,Kudler-Flam:2025cki,Sasieta:2025vck,Belin:2025ako,VanRaamsdonk:2026tnv,Maldacena:2024uhs,Ivo:2024ill,maldacena_real_2025,Chen:2025jqm,usatyuk_closed_2024,Usatyuk:2024isz,abdalla_gravitational_2025,harlow_quantum_2025,Akers:2025ahe,Blommaert:2025bgd,Nomura:2025whc,Chen:2025fwp}.

In 1983, Hartle and Hawking~\cite{hartle_wave_1983} proposed that the state $\ket \Psi$ of our universe is the state with no boundaries at all:\footnote{The Hartle-Hawking proposal assumes that our universe is spatially closed. This does not conflict with observation, which only yields a lower bound on the size of the universe. Applications of Eq.~\eqref{eq:hh} have mostly been restricted to a single universe, i.e., the case where $J$ is connected. This restriction is not necessary, as stressed in Sec. VIII of the original paper. However, no proposal for the relative normalization of $J$'s with multiple connected components was made.} 
\begin{equation}\label{eq:hh}
       \ket\Psi=\ket\varnothing~.
\end{equation}
By Eq.~\eqref{eq:ip} and the Born rule, the Hartle-Hawking proposal \eqref{eq:hh} implies that the probability for observing a universe with data $J$ is given by
\begin{equation}\label{eq:probdef}
    P(\varnothing\to J) =  \frac{|\braket{J}{\varnothing}|^2}{\braket{J}{J}\braket{\varnothing}{\varnothing}} = \frac{|G(J)|^2}{G(J^*,J)G(\varnothing)}~.
\end{equation}
In this paper, we will adopt this proposal and explore its implications. We find that when $J$ is a single connected universe, 
\begin{equation}\label{eq:hhprob}
    P(\varnothing\to J) \approx 1~.
\end{equation}
In other words, the states $\ket J$ of a closed universe are all nearly parallel to the Hartle-Hawking state. 

Note that Eq.~\eqref{eq:probdef} assumes that the state $\ket{J}$ is normalizable, which, as we shall see later in the paper, is not the case for closed universe states with fixed metric or fixed extrinsic curvature. To compute probabilities for such states, we need to define a smeared projector. We shall show 
that our result \eqref{eq:hhprob} holds when the state and its smearing are both in the semiclassical regime.

Our result deviates drastically from Ref.~\cite{hartle_wave_1983} and upends the extant literature on the Hartle-Hawking proposal. (For an excellent early review, see Ref.~\cite{Halliwell:1989myn}; see \cite{lehners_review_2023,Maldacena:2024uhs} for recent reviews and references.) The discrepancy can be traced to the following two assumptions implicit in \cite{hartle_wave_1983}, which our explicit calculations show to be false:
\begin{enumerate}
    \item Hartle and Hawking implicitly assumed that the set of states specified by all three-metrics and fields, as in Eq.~\eqref{eq:jh}, are mutually orthogonal, so that the normalized inner product
    \begin{equation}\label{eq:hh3}
        \Psi(J)\equiv \frac{\braket{J}{\varnothing}}{\braket{J}{J}^{1/2}\braket{\varnothing}{\varnothing}^{1/2}}
    \end{equation}
    admits an interpretation as a wavefunction. (More precisely, certain states are approximately gauge equivalent, for example if the corresponding data $J$ are related by classical Lorentzian evolution. Ref.~\cite{hartle_wave_1983} implicitly assumed that it suffices to quotient by classical Lorentzian evolution.) An analogous assumption was also made for states specified by the extrinsic curvature.
    \item Hartle and Hawking assumed that the norm $\braket{J}{J}$ in the denominator of Eq.~\eqref{eq:hh3} does not depend on $J$. This is implicit in Eq.~(3.1) of \cite{hartle_wave_1983}, which posits that the gravitational path integral with the single boundary $J$ computes the wavefunction of our universe, up to a $J$-independent normalization factor $N$:
    \begin{equation}\label{eq:hhprobwrong}
    \Psi(J)\equiv \frac{\braket{J}{\varnothing}}{\braket{J}{J}^{1/2}\braket{\varnothing}{\varnothing}^{1/2}}~~ \stackrel{?}{=} N\, G(J)  = N\times\left(\inlinefig[4]{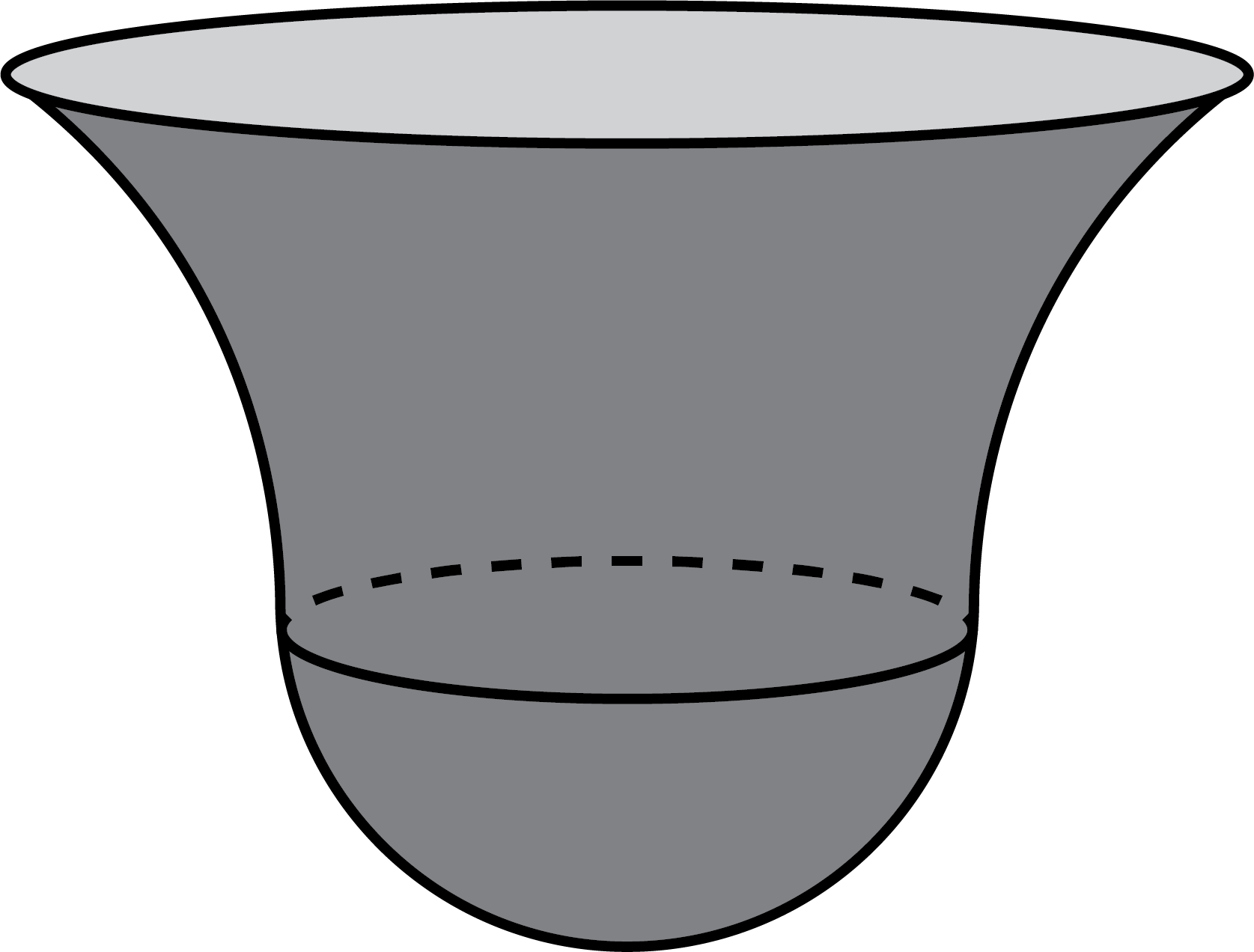}\right)~.
    \end{equation}
    By the Born rule, this yields a probability
    \begin{equation}
    P(\varnothing \to J) = |\Psi(J)|^2 ~~\stackrel{?}{\propto} |G(J)|^2
    \label{eq:old-probability-HH}
    \end{equation}
    for finding the universe in the state $J$. (The question marks indicate relations that prove false upon explicit calculation.)
\end{enumerate} 
Explicit calculation reveals, however, that
\begin{equation}\label{eq:wtaf}
    G(J)\, G(J^*) \approx G(J^*,J) \, G(\varnothing)\quad \Rightarrow \quad P(\varnothing \to J) \approx  \frac{\left|\inlinefig[3]{Figures/Kphi_HH.png}\right|^2}{\left|\inlinefig[3]{Figures/Kphi_HH.png}\right|^2+\inlinefig[4]{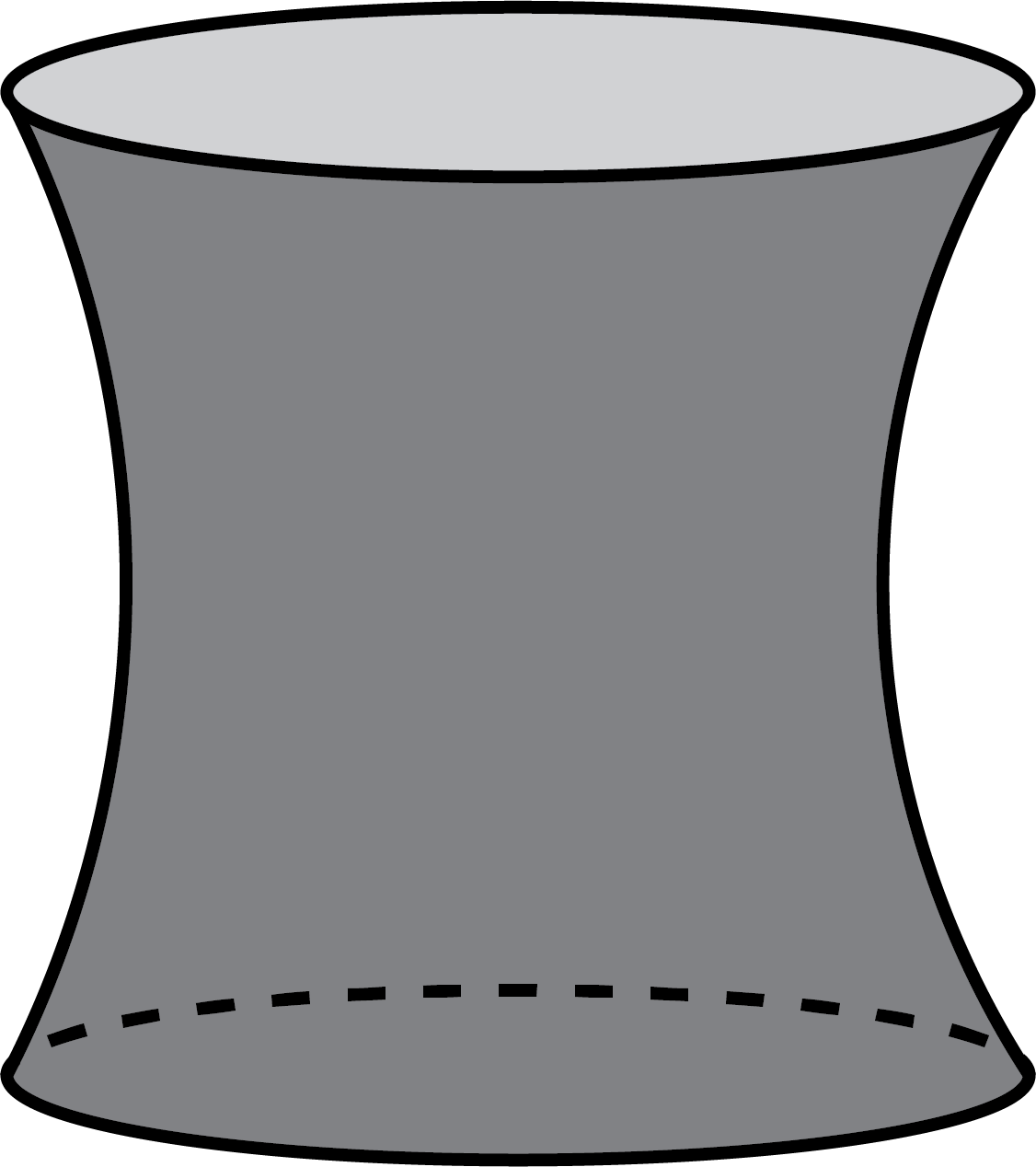}}\approx 1\,.
\end{equation}
This is because at leading order in the semiclassical expansion ($G_N \to 0$), the same geometry dominates the left and the right-hand side. This saddlepoint consists of two disconnected spacetimes with boundaries carrying the data $J$ and $J^*$, respectively. Nonperturbative corrections to this result come from geometries that connect $J$ to $J^*$. In cosmological settings to which the Hartle-Hawking proposal is widely applied (and in many other settings), these connected corrections are in fact suppressed by factors of the form $\exp(-L^2/G_N)$. Hence, they are negligible in the semiclassical limit where the characteristic curvature length $L$ is much larger than the Planck scale. By Eq.~\eqref{eq:wtaf}, Eq.~\eqref{eq:probdef} evaluates to Eq.~\eqref{eq:hhprob}, and not to Eq.~\eqref{eq:old-probability-HH}. This implies that the states $\ket J$ are nearly parallel, invalidating the interpretation of $\Psi(J)$ as a wavefunction of its argument.

\paragraph{Outline} 

In Sec.~\ref{sec:inner-prod-review}, we briefly review the gravitational path integral and its interpretation as an inner product. We also introduce a second, ``statistical'' interpretation, which has gained prominence over the recent decade \cite{Saad:2018bqo,Penington:2019kki,saad_jt_2019,Bousso:2019ykv,marolf_transcending_2020,Sasieta:2022ksu,Balasubramanian:2022gmo,Chandra:2022bqq,Chandra:2023dgq,DiUbaldo:2023qli,Jafferis:2025jle,marolf_nature_2024,usatyuk_closed_2024,Usatyuk:2024isz,abdalla_gravitational_2025,harlow_quantum_2025}. In this approach, a single GPI $G(J_1, \ldots, J_n)$ is interpreted as computing an average over an ensemble of inner product, not only for a single amplitude, but also for any product of amplitudes involving the connected components $J_i$ in its argument.

In Sec.~\ref{sec:nonpert-prob}, we compute the probability for finding the data $J$ in the Hartle-Hawking state $\ket\varnothing$. With the conventional interpretation of the GPI, we find Eq.~\eqref{eq:hhprob}. With the statistical interpretation, we find that the probability for any $J$ is exactly 1 in every member of the ensemble of theories. This is consistent with recent general arguments that the Hilbert space of closed universes in each member of the ensemble is one-dimensional~\cite{Penington:2019npb, McNamara:2020uza, marolf_transcending_2020, blommaert_gravity_2022, blommaert_alpha_2022,  usatyuk_closed_2024, Usatyuk:2024isz, abdalla_gravitational_2025}.

In Sec.~\ref{sec:inf}, we illustrate our results in the context of slow-roll cosmic inflation. We first evaluate the Hartle-Hawking proposal according to the traditional (inconsistent) Eq.~\eqref{eq:hhprobwrong}, reproducing the infamous prediction of an empty universe~\cite{Vilenkin:1982de,linde1984,Vilenkin:1987kf,Maldacena:2024uhs}. We then exhibit the most relevant saddlepoints that contribute to Eq.~\eqref{eq:hhprob}.

In Sec.~\ref{sec:discussion}, we discuss the implications of our result for the quantum description of closed universe states. A key challenge is to associate distinct data to orthogonal states. Ironically, this requires projecting out the no-boundary state after associating data to quantum states. We also discuss how the statistical approach can be obtained from the Hilbert space implied by the conventional approach to the GPI, using the orthonormal ``$\alpha$-basis'' constructed by Marolf and Maxfield~\cite{marolf_transcending_2020}.

\section{Inner Products From the Gravitational Path Integral}
\label{sec:inner-prod-review}

In field theory, the path integral computes the unnormalized inner product between state vectors defined by the boundary conditions imposed at specific times on a fixed background manifold: 
\begin{equation}
    \langle\phi_1,\tau_1|\phi_2,\tau_2\rangle = \langle\phi_1|e^{-H(\tau_1-\tau_2)}|\phi_2\rangle = \int _{\phi(\tau_1)=\phi_1}^{\phi(\tau_2)=\phi_2}\cD \phi \exp\left(-I_{QFT}[\phi]\right)\,.
    \label{eq:qftPI}
\end{equation}

In a theory with gravity, the Hamiltonian $H$ becomes a constraint and vanishes for closed universes \cite{Arnowitt:1962hi,dewitt}. Moreover, the field configurations we sum over include the spacetime metric, so the path integral $G(J)$ is over all geometries with boundary $J$. This leads to Eq.~\eqref{eq:definition-G}. 

State vectors $\ket J$ are defined in terms of boundary conditions $J$ that include a three-metric or extrinsic curvature, as described in the introduction.\footnote{ We work in the co-invariant formalism \cite{hartle_wave_1983,Marolf:2000iq,Held:2025mai,Banihashemi:2024aal}. This is the appropriate choice when computing inner products using the GPI. The GPI automatically imposes the gravitational constraints when computing inner products. } Here we consider only closed universes; when $J$ has more than one connected component, we write $\ket{J} = \ket{J_1,\ldots,J_n}$.  As in QFT, the associated dual vector $\bra{J_1...J_n}$ is obtained by an antilinear operation $\mathcal T$ that complex conjugates the boundary conditions and, for oriented boundaries, reverses the orientation of the boundary manifolds $\Sigma_i$ (see for instance \cite{witten_bras_2025}).  The Hartle-Hawking state $\ket\varnothing$ is the state with no boundary at all \cite{hartle_wave_1983}.

\paragraph{Conventional approach} In quantum cosmology,  the GPI has been interpreted in strict analogy to Eq.~\eqref{eq:qftPI}, as computing an unnormalized inner product. For example, if $J$ consists of two connected components $J_1$, $J_2$, then $G(J_1,J_2)$ is the transition amplitude in a closed universe $J_1^*$ to a closed universe with data $J_2$. 

The conventional interpretation of the GPI contains features that are absent in theories without gravity. Because there is no preferred time direction, $G(J_1,J_2)$ also admits an interpretation as a transition from $J_2^*$ to $J_1$, or as the creation of two universes from nothing:
\begin{equation}\label{eq:multi}
    G(J_1,J_2) = \braket{J_1^*}{J_2} = \braket{J_2^*}{J_1}
    =\braket{J_1^*,J_2^*}{\varnothing} = \ldots~.
\end{equation}
By the Born rule, the normalized probability for the transition is\footnote{If $\ket{J_1}$ and $\ket{J_2}$ are related by the action of the Hamiltonian constraint (i.e., they lie on the same gauge orbit), then the GPI implies that $P(\psi \to J_1) = P(\psi\to J_2)$ for any state $\ket{\psi}$, and hence $P(J_1\to J_2)=1$. This should not be confused with the findings of this paper, in which many transition probabilities are found to be either close to 1 or exactly 1 for reasons unrelated to the Hamiltonian constraint.}
\begin{equation}\label{eq:prob12}
    P(J_1\to J_2) = \frac{|G(J_1,J_2)|^2}{G(J_1^*,J_1)\,G(J_2^*,J_2)}~.
\end{equation}
It is not obvious that the GPI leads to a positive norm, e.g., that $G(J^*,J)\geq 0$ for all $J$. Here we follow the common practice of assuming this.

The norm of the Hartle-Hawking state consists of all possible spacetimes with no boundaries; they need not be connected:
\begin{equation}
    \mathcal{N}= \exp{\inlinefig[4]{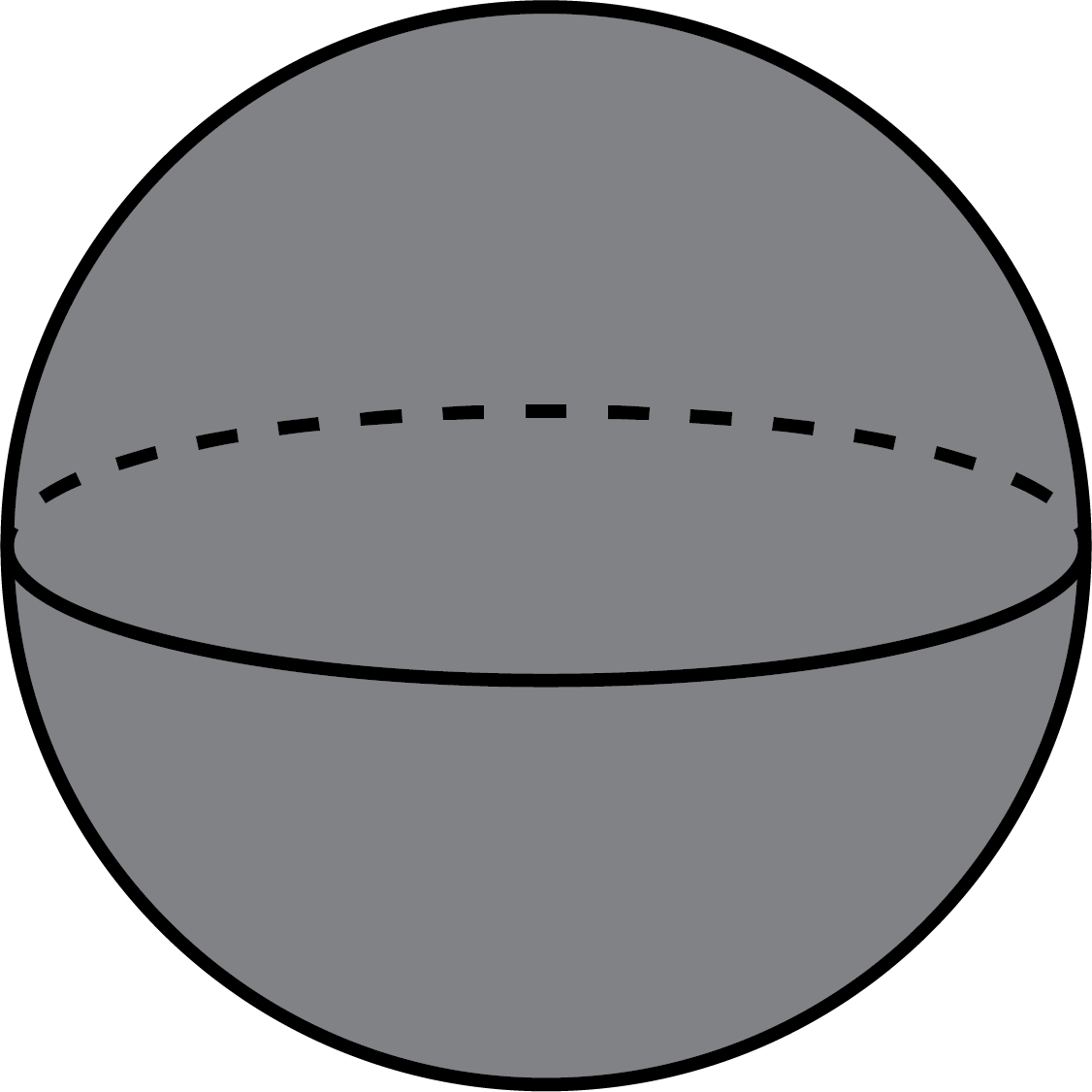}+\inlinefig[4]{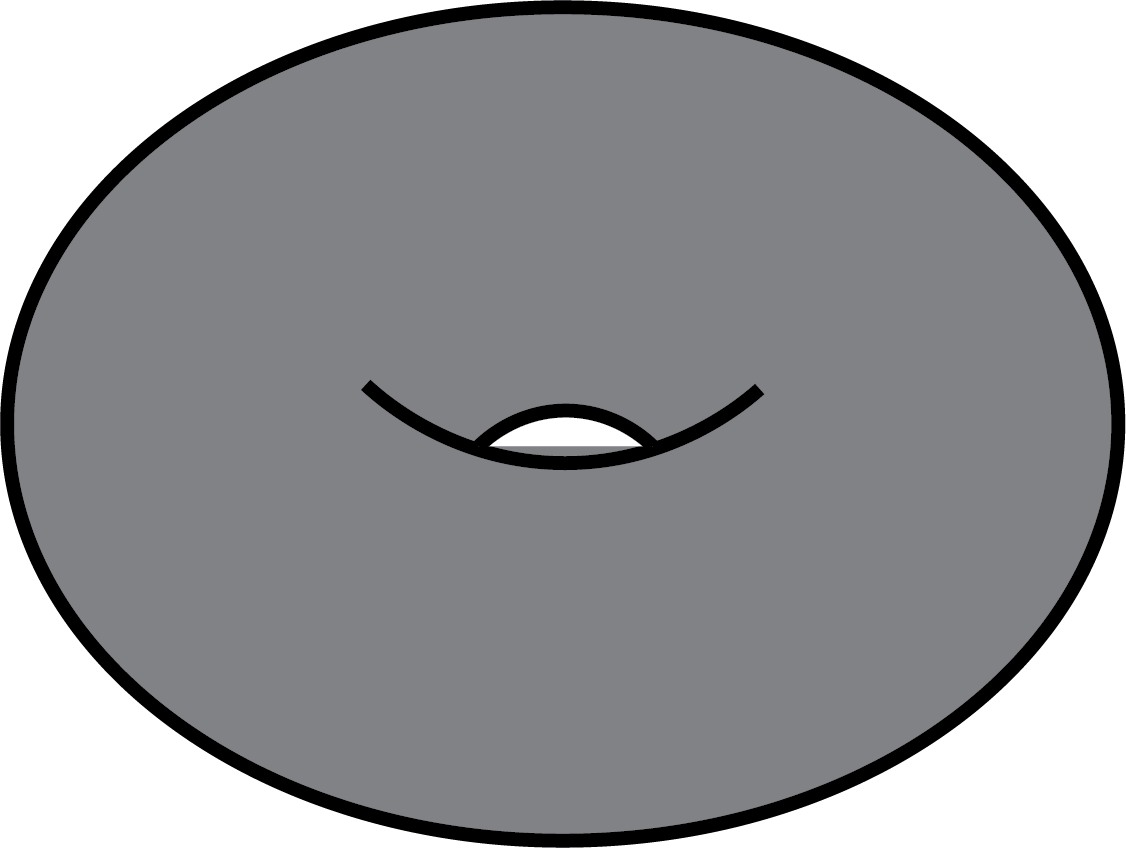}+\cdots}\,,
    \label{eq:vacuumN}
\end{equation}
This factor contributes to every $G(J)$. It cancels in normalized quantities such as Eq.~\eqref{eq:prob12}.

\paragraph{Statistical approach}
In recent years, a different interpretation of the GPI has become increasingly prevalent \cite{Saad:2018bqo,Penington:2019kki,saad_jt_2019,Bousso:2019ykv,marolf_transcending_2020,Sasieta:2022ksu,Balasubramanian:2022gmo,Chandra:2022bqq,Chandra:2023dgq,DiUbaldo:2023qli,Jafferis:2025jle,marolf_nature_2024,usatyuk_closed_2024,Usatyuk:2024isz,abdalla_gravitational_2025,harlow_quantum_2025}. To motivate this approach, consider the following qualitative difference between path integrals with and without gravity. Field theory amplitudes factorize when computed on different background spacetimes $A,B$:
\begin{equation}
    \braket{\phi^A_1,\phi^B_2,\tau^A_1,\tau^B_2}
    {\phi^A_3,\phi^B_4,\tau^A_3,\tau^B_4}=
    \braket{\phi^A_1,\tau^A_1}{\phi^A_3,\tau^A_3}
    \braket{\phi^B_2,\tau^B_2}{\phi^B_4,\tau^B_4}~.
\end{equation}
Here, the left-hand side is evaluated on the manifold $A\sqcup B$, and the two factors on the right-hand side are computed, respectively, on $A$ only and on $B$ only.

By contrast, amplitudes computed from the GPI do not factorize \cite{Maldacena_2004}. For example, 
\begin{align}\label{eq:nonfactorize}
    G(J_1^*, J_2^*, J_3, J_4) &\neq G(J_1^*, J_3) G(J_2^*, J_4)~, 
\end{align}
because the left-hand side contains contributions from connected geometries such as\footnote{
    Notice that this inequality is not only a consequence of mismatching powers of $\cN$. Here, we emphasize that, even after discounting closed disconnected spacetimes, equality does not hold as seen in \eqref{eq:example-ensemble}. 
}
\be 
    G(J_1^*, J_2^*, J_3, J_4) &\supset \inlinefig[6]{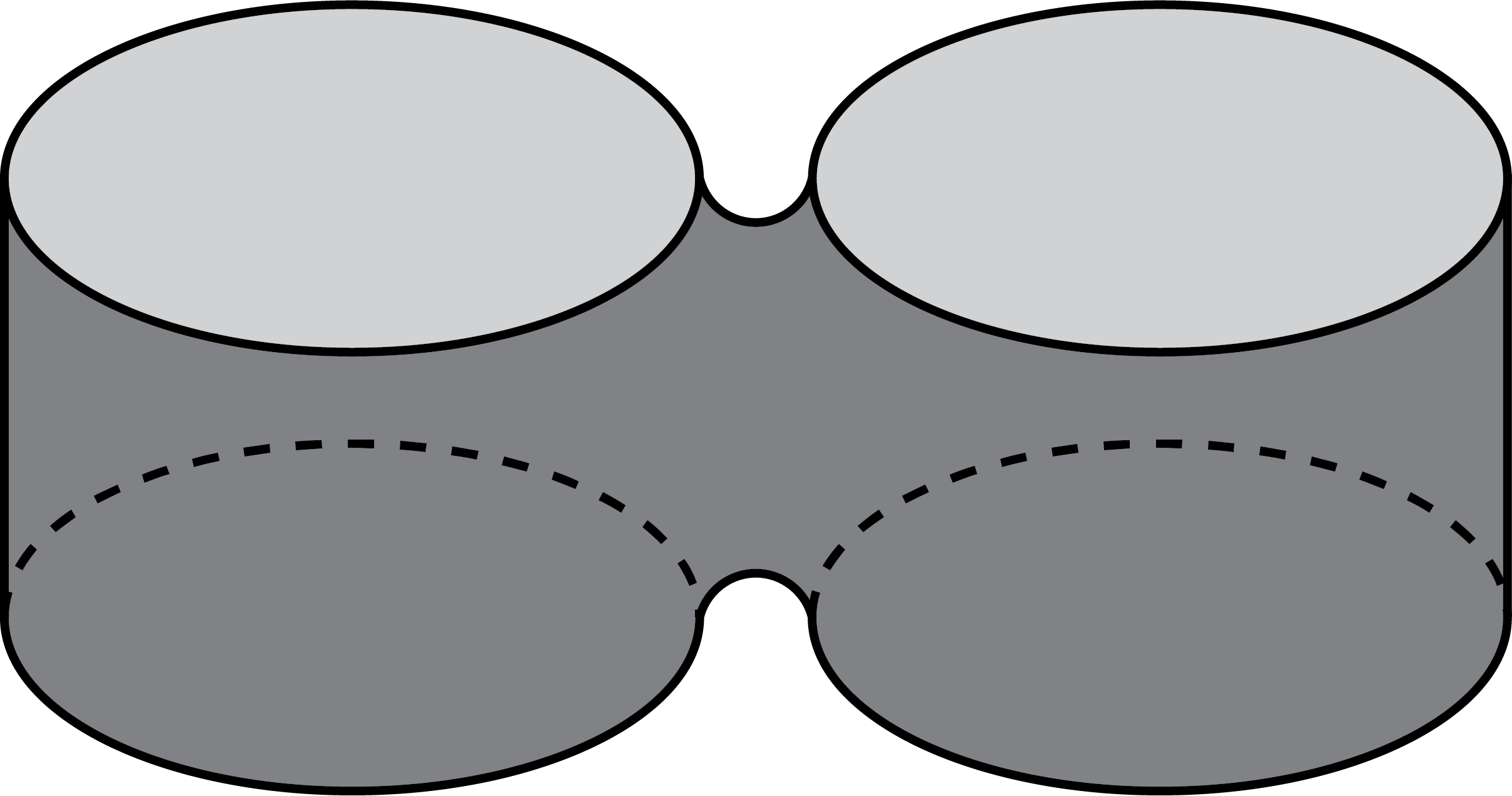} \not\subset G(J_1^*, J_3) G(J_2^*, J_4)\,.
\ee
In the conventional approach, the left-hand-side has no relevance for computing a product of amplitudes. Nevertheless, can a single path-integral, such as $G(J^*_1, J^*_2, J_3, J_4) $, still capture properties of products of inner products?
This is possible by adopting an alternate interpretation of the GPI where $G(J_1^*,J_2^*, J_3, J_4)$ is viewed as computing a probability-weighted average over a presumed \emph{ensemble} of inner products. Denoting an average over this ensemble by an overbar, we have
\begin{align}
    \mathcal{N}^{-1} G(J_1^*, J_2^*, J_3, J_4) = \overline{\braket{J_1}{J_3}\braket{J_2}{J_4}} \neq \overline{\braket{J_1}{J_3}}\times \overline{\braket{J_2}{J_4}} = \mathcal{N}^{-2} G(J_1^*, J_3) G(J_2^*, J_4)\,.
    \label{eq:example-ensemble}
\end{align}
The factors of $\mathcal N$ (defined in \eqref{eq:vacuumN}) ensure that the averages over the ensemble come from a normalized probability distribution. 

Generalizing to an arbitrary number of boundary conditions, 
the statistical interpretation of the GPI states that the $G(J_1,\ldots,J_n)$ computes not only the ensemble average of any single inner product that can be written by distributing its arguments $J_i$ between one bra and one ket, as in the conventional interpretation above; see Eq.~\eqref{eq:multi}. In the statistical interpretation, $G(J_1,\ldots,J_n)$ can \emph{also} be used to compute the ensemble average of any \emph{product} of inner products that can be constructed from its arguments. For example,
\begin{align}
    \mathcal N^{-1}G(J_1^*, \dots, J_n^*, \tilde J_1, \dots, \tilde J_{ n}) &=
   \overline{\langle J_1 \dots J_n  |\tilde J_1\dots  \tilde J_n\rangle} \\ &=
   \overline{\langle J_1|\tilde J_1\rangle
   \langle J_2 \dots J_n  |\tilde J_2\dots  \tilde J_n\rangle} = \ldots \\ &=
   \overline{\langle J_1|\tilde J_1\rangle \dots \langle J_n|\tilde J_n\rangle}  \,.
 \label{eq:the-two-approaches-1}
\end{align} 
The $\ldots$ include all possible products of inner products that can be formed by distributing the arguments of $G$ into an arbitrary number of bra-ket pairs. 
The non-factorization property of $G$, in this approach, arises from the fact that averaging and taking products need not commute. By computing appropriate moments, properties of the underlying statistical ensemble can be extracted from the GPI.

The statistical approach has proved particularly useful in resolving puzzling features of the GPI in the context of holography \cite{tHooft:1993dmi,Susskind:1994vu,Maldacena:1997re,Witten:1998qj,Gubser:1998bc,Aharony:1999ti}. Suppose, for example, that two boundary conditions $J_1 = J_2$ are of the form $S_1\times S_2$. In a theory with a nonpositive cosmological constant, $G(J_i)$ has the interpretation of computing the thermal partition function of a quantum-mechanical theory without gravity. This was anticipated by Gibbons and Hawking~\cite{Gibbons:1976ue}. The AdS/CFT correspondence explicitly identifies some of the relevant quantum-mechanical theories~\cite{Maldacena:1997re,Maldacena:1998bw,Aharony:2008ug,Maldacena:2016upp,kitaev,Stanford:2017thb,saad_jt_2019}. Since such theories do not contain gravity, their partition functions should factorize, in apparent conflict with Eq.~\eqref{eq:nonfactorize}~\cite{Maldacena_2004}. A closely related puzzle arises~\cite{Bousso:2019ykv} from the recent GPI computation~\cite{Penington:2019kki,Almheiri:2019qdq} of the entropy of the Hawking radiation emitted by a black hole. 

In versions of AdS/CFT in two bulk spacetime dimensions, explicit ensembles have been found~\cite{saad_jt_2019,Ghosh:2019rcj,Witten:2020wvy,Saad:2019pqd,Cotler:2016fpe,Stanford:2017thb}. In three-dimensional models, this is an area of active research~\cite{Chandra:2022bqq,DiUbaldo:2023qli,Jafferis:2025jle}. 
So far, no explicit ensemble has been identified 
in four or higher dimensions. Nevertheless, even in such examples, the gravitational contributions that lead to the factorization puzzle provide useful information about the statistics of CFT data~\cite{Sasieta:2022ksu,Balasubramanian:2022gmo,Balasubramanian:2022lnw,deBoer:2023vsm}.

\section{Probabilities in the Hartle-Hawking State}
\label{sec:nonpert-prob}

In this section, we use the gravitational path integral to compute the probability for a closed universe $\ket J$ in the Hartle-Hawking state $\ket{\varnothing}$. In the conventional approach, we will find that the probability for $\ket J$ is 1 up to nonperturbative corrections.  Using the statistical approach, we will find that the probability is 1 in every member of the ensemble of theories. 

Thus, the Hartle-Hawking state is parallel, or at least nearly parallel, to every simple geometric state of a closed universe, and no useful predictions can be made. This is unsurprising in the statistical approach, which has already been shown to imply that the Hilbert space of closed universes is one-dimensional~\cite{Penington:2019npb,marolf_transcending_2020,usatyuk_closed_2024,Usatyuk:2024isz,abdalla_gravitational_2025,harlow_quantum_2025}, so all inner products of normalized states must be a pure phase. In the conventional approach, the Hartle-Hawking proposal had a fighting chance because the Hilbert space is nontrivial. Our result, however, shows that the proposal does not discriminate usefully between naively different closed universes.

\paragraph{Hartle-Hawking probabilities in the conventional approach}
\label{sec:HHMM}

In the conventional interpretation of the GPI, one computes a (single) normalized probability for the universe with data $J$, in the Hartle-Hawking state: 
\begin{align}
    P(\varnothing\to J) =\frac{|\braket{J}{\varnothing}|^2}{\braket{J}{J} \braket{\varnothing}{\varnothing}} =\frac{|G(J)|^2}{G(J^*, J)\, \mathcal{N}}~.
\end{align}
Each inner product must be evaluated as a separate GPI. In the small $G_N$ expansion, the numerator and denominator include saddlepoint contributions of the form
\begin{align}
     |G(J)|^2 &=\mathcal{N}^2 \left|\inlinefig[3]{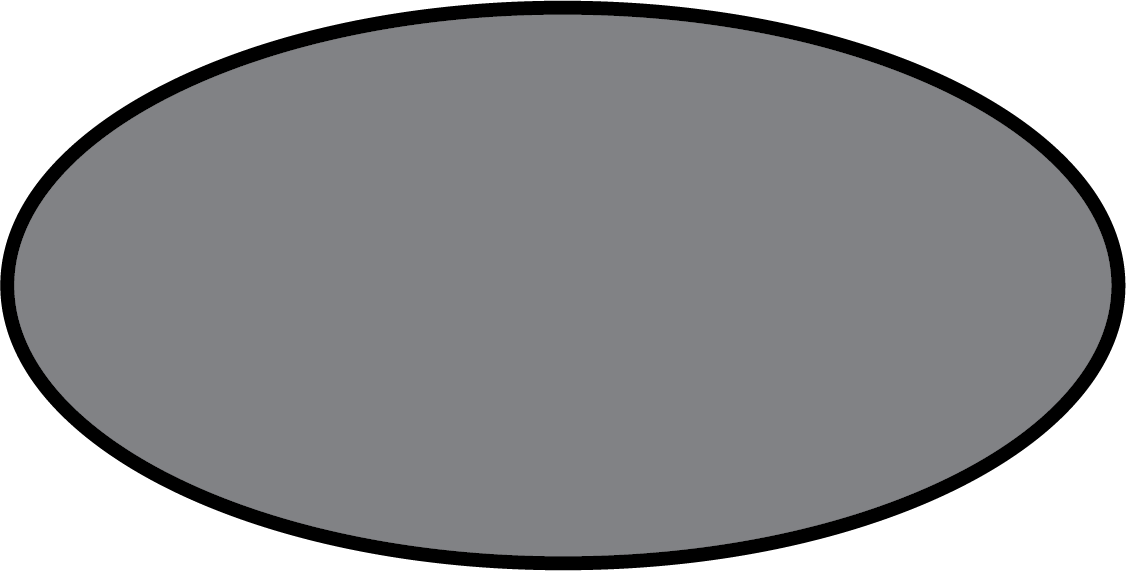}+\inlinefig[3]{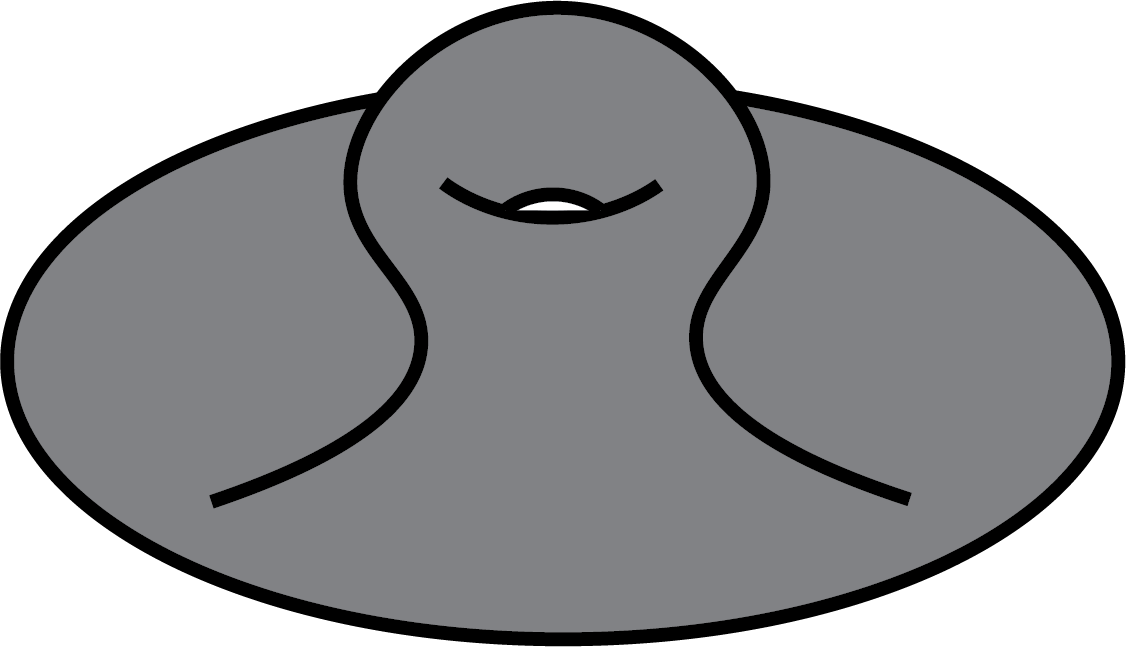}+\cdots\right|^2~,\label{num}\\
G(J^*,J)\, \mathcal{N}  & = \mathcal{N}^2 \left(\inlinefig[4]{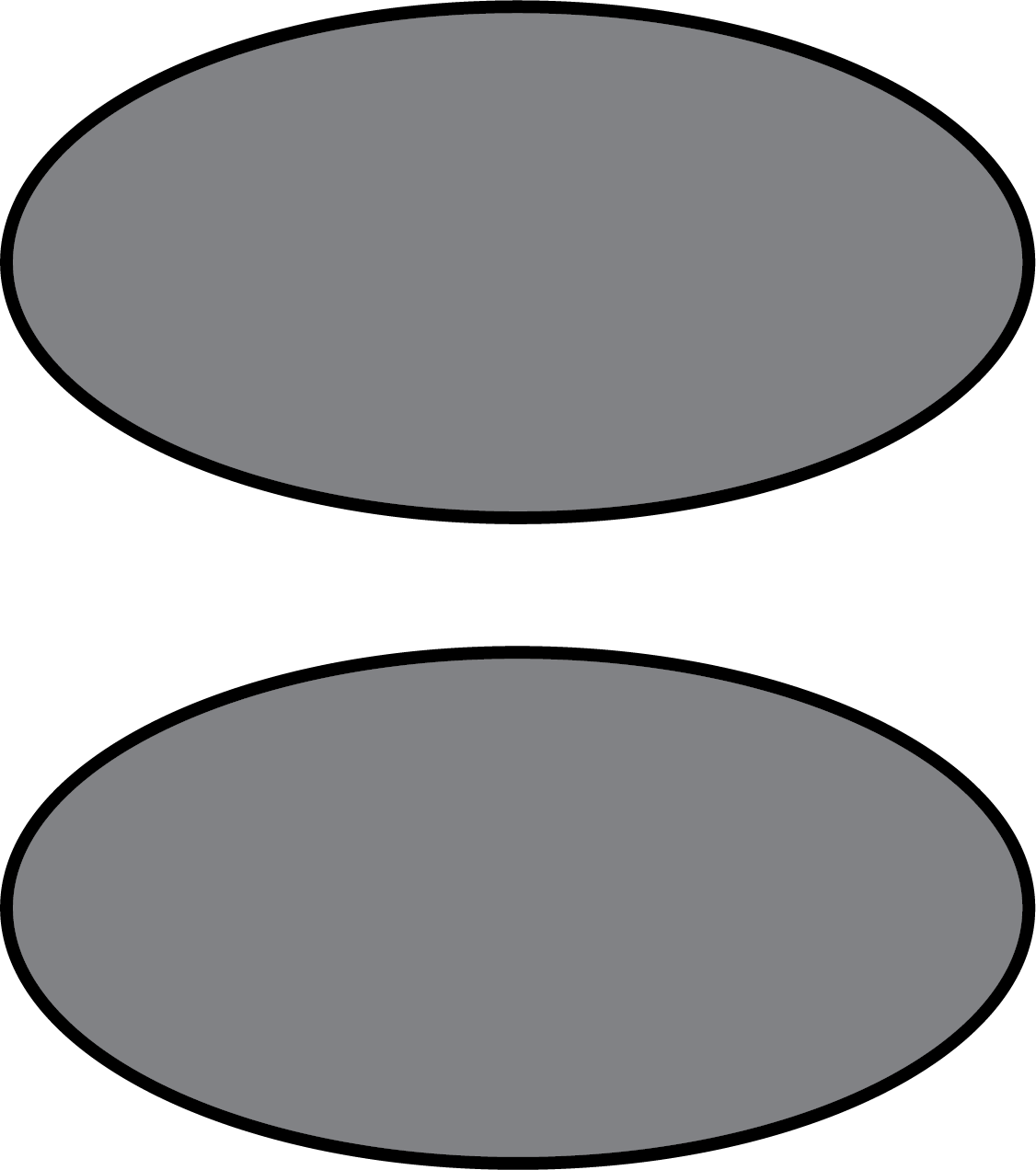}+\inlinefig[4]{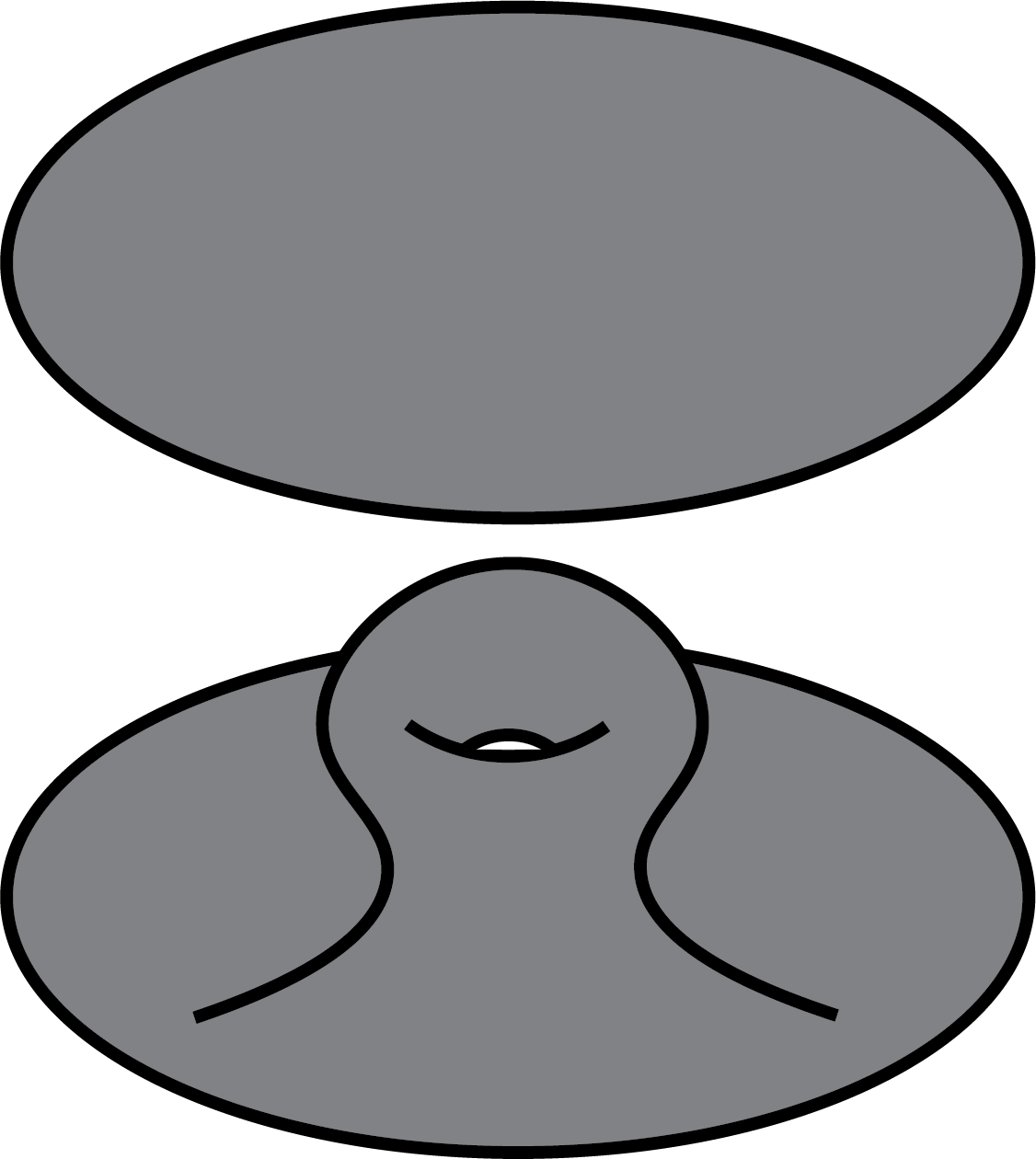}+\inlinefig[4]{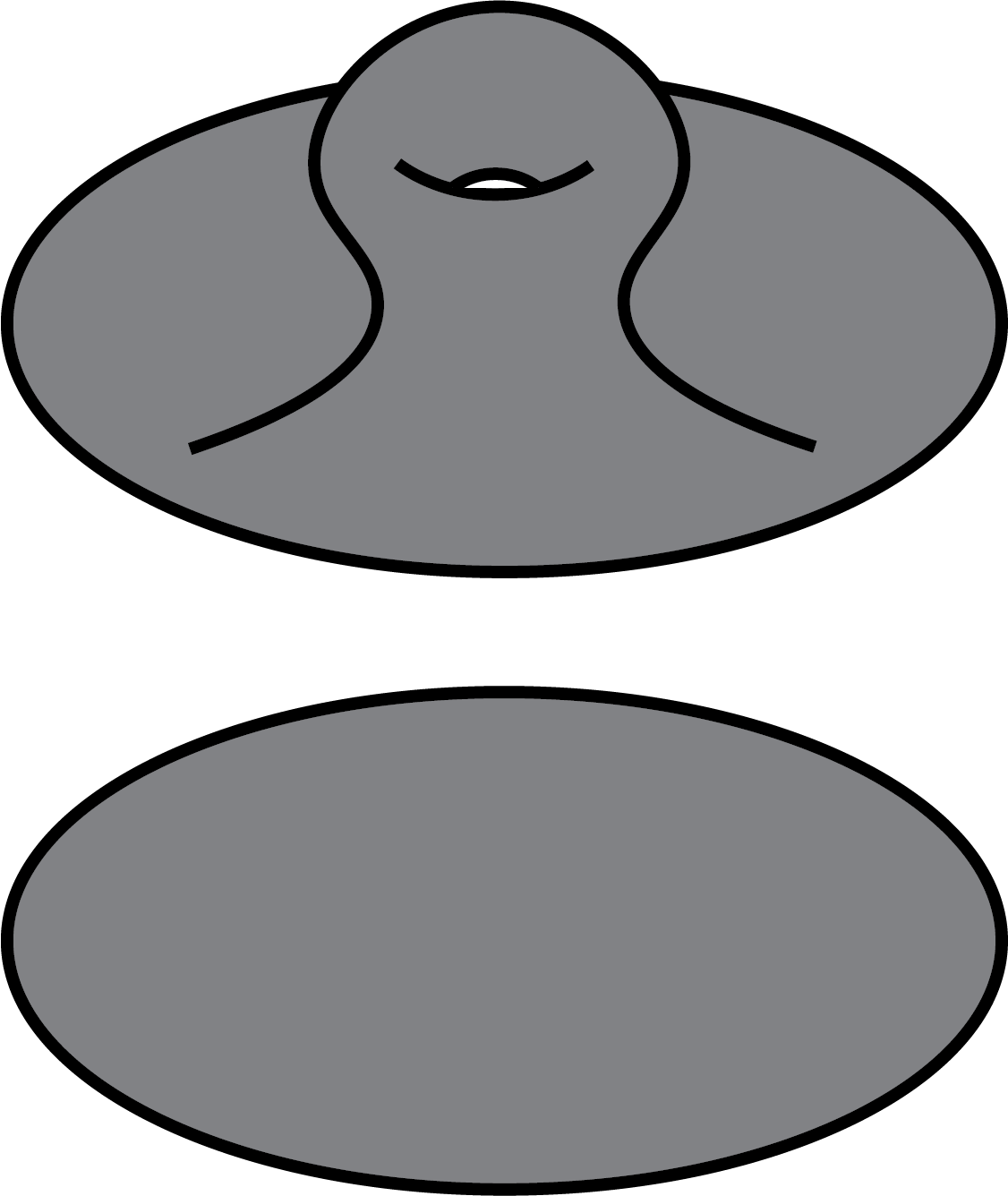}+\cdots+\inlinefig[4]{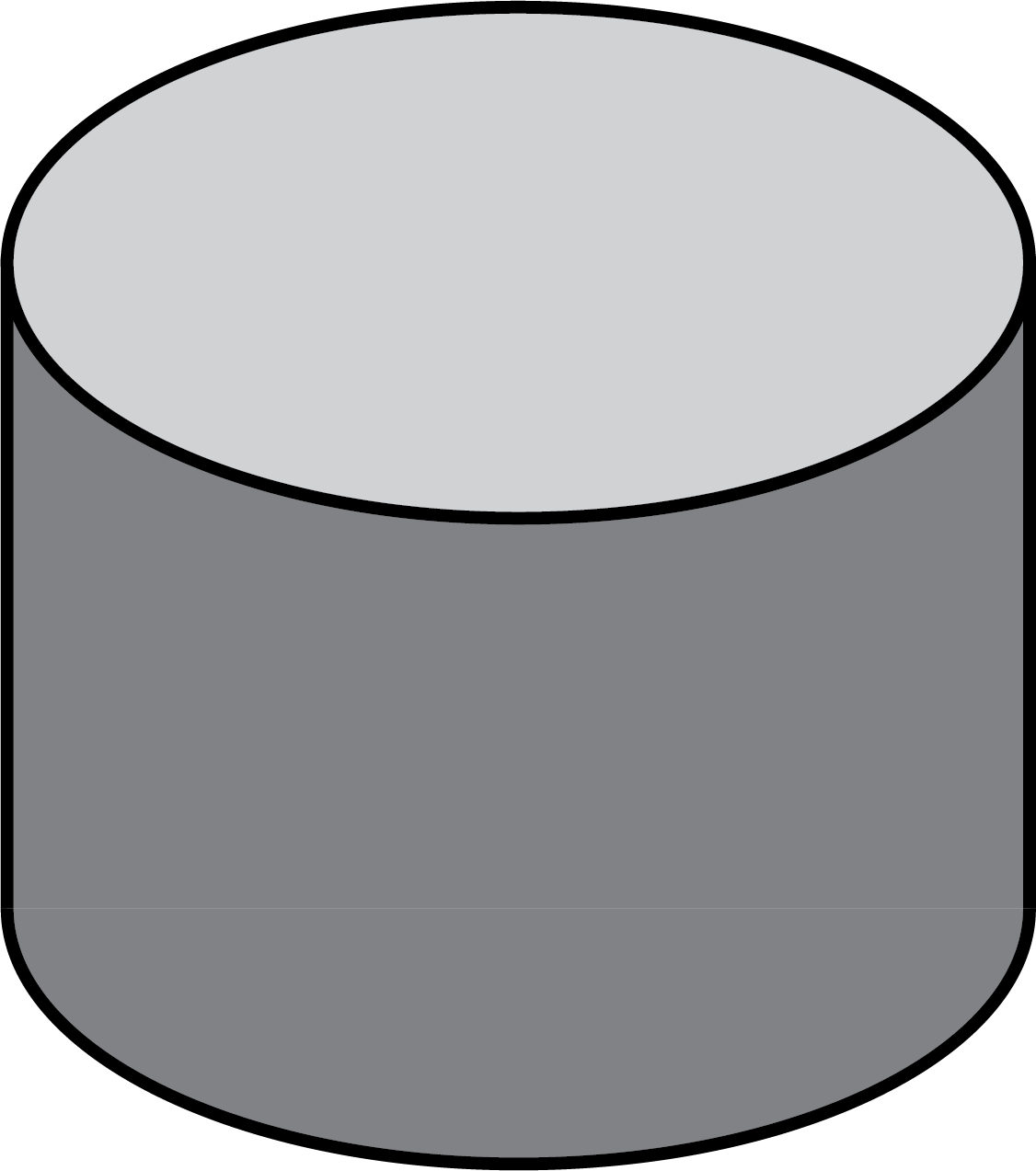}+\cdots\right),\label{den}
\end{align}
where we have schematically represented the leading saddle by a disk and subleading saddles by higher genus geometries. Here, $L$ is the smallest characteristic length scale for the saddlepoints. 
Note that Eq.~\eqref{den} contains all of the terms that appear in Eq.~\eqref{num}. In addition, Eq.~\eqref{den} contains geometries that connect the two boundaries. Thus, 
\begin{align}\label{eq:main}
    P(\varnothing\to J) = \left(1+\frac{\inlinefig[4]{Figures/02.png}+\inlinefig[4]{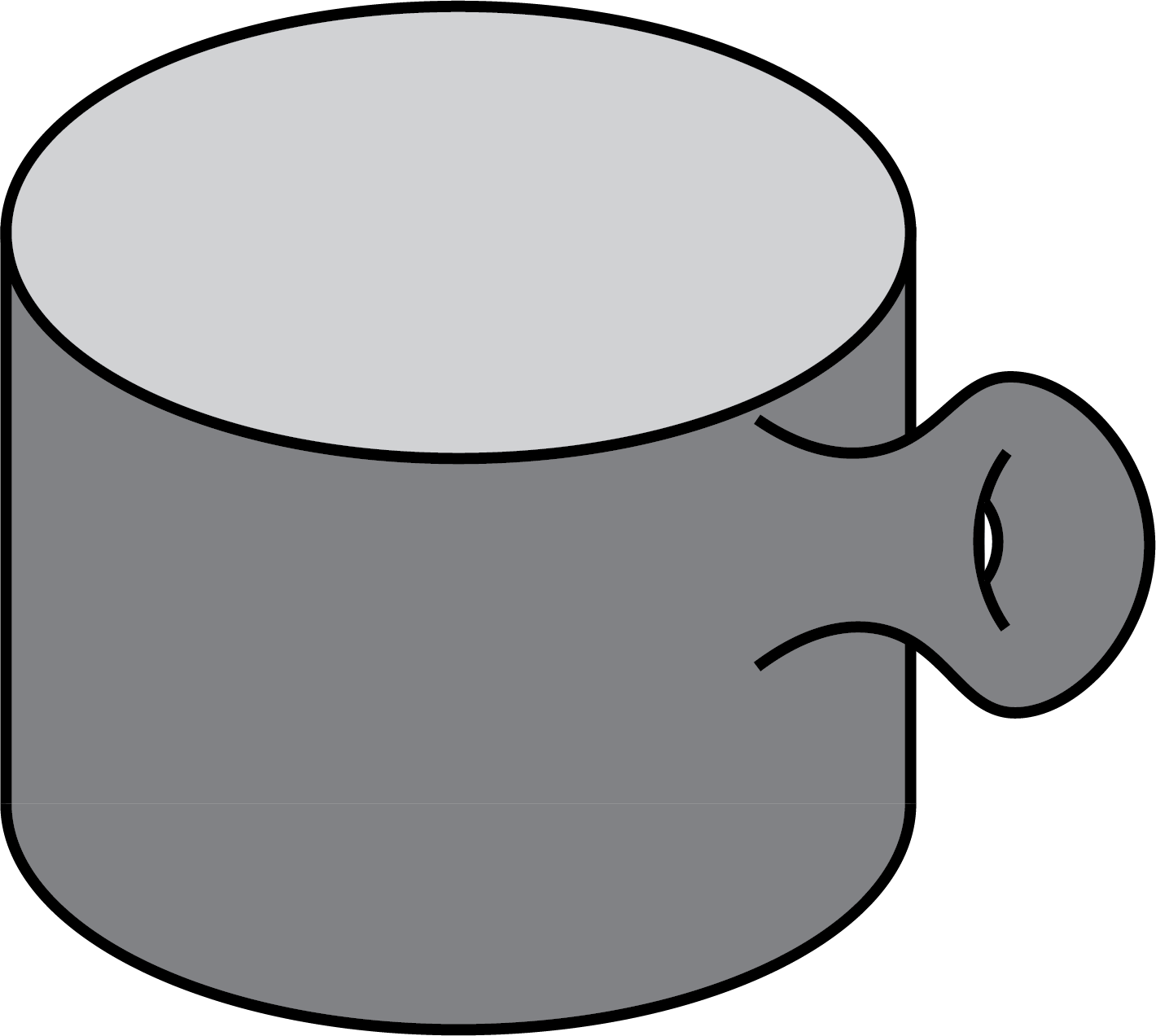}+\cdots}{\inlinefig[4]{Figures/0101.png}+\inlinefig[4]{Figures/0111.png}+\inlinefig[4]{Figures/1101.png}+\cdots}\right)^{-1}= \frac{1}{1 + \frac{\text{connected}}{\text{disconnected}}}\quad .
\end{align}
In many known examples, including the cosmological setting of interest here, connected contributions in the numerator are suppressed by $\exp(-L^{d-2}/G_N)$, relative to the disconnected geometries in the denominator.
This is true both in the saddle point approximation for theories in AdS,\footnote{We are aware of two noteworthy exceptions. The first is when the gravitational EFT has an exact global symmetry. In such cases, states defined by boundary conditions $J$ that carry a global symmetry charge are exactly orthogonal to the no-boundary state. Such global symmetries are, however, expected to be broken in UV complete theories of quantum gravity \cite{Misner:1957mt, Giddings:1988cx, Kallosh:1995hi, Polchinski:2003bq, ArkaniHamed:2006dz, Banks:2010zn, Harlow:2018jwu, Harlow:2018tng, Harlow:2020bee, Chen:2020ojn, hsin_violation_2020}. The second exception is discussed in \cite{Maldacena_2004} in the construction of meron solutions in AdS$_4$. There, the two-boundary connected saddle was found to dominate over the disconnected saddle for sufficiently small values of the Yang-Mills coupling. Nevertheless, just as with the case of global symmetries, it is unclear whether such values of the Yang-Mills coupling can arise in UV complete theories of quantum gravity. Even if one believes in the validity of such gravitational EFTs, the evaluation of the no-boundary proposal completely differs from the original treatment of \cite{hartle_wave_1983}.  } and when including off-shell geometries for theories, like JT gravity, in which we have full control over the GPI \cite{saad_jt_2019}. As an example, in Sec. \ref{sec:inf} we will verify this for an inflationary model when boundary conditions are smeared to define normalizable states.  
Thus, in the conventional approach, the probabilities for simple geometries $\ket{J}$ approach unity in the $G_N\to 0$ limit. However, they are not exactly 1, consistent with the fact that the Hilbert space is nontrivial due to the non-factorization of $G$.

\paragraph{Hartle-Hawking probabilities in the statistical approach}
\label{sec:HHstat}

In the statistical approach, $|\langle J|\varnothing\rangle |^{2}$ and $\langle J|J\rangle\langle \varnothing|\varnothing\rangle$ are random variables in an ensemble of theories. Consider the GPI with $2n$ connected boundaries, $n$ of which have the boundary condition $J$, and $n$ of which have boundary condition $J^*$. For all $n\geq 1$, 
\begin{equation}
    G[J,\ldots,J,J^*,\ldots, J^*] = \overline{|\langle J|\varnothing\rangle |^{2n}}=\overline{\left(\langle J|J\rangle\langle \varnothing|\varnothing\rangle\right)^n}
\end{equation}
by Eq.~\eqref{eq:the-two-approaches-1}. Thus, $|\langle J|\varnothing\rangle |^{2}$ and $\langle J|J\rangle\langle \varnothing|\varnothing\rangle$ furnish identical solutions to the moment problem. Therefore, the probability distributions over these variables in the ensemble are identical: up to a measure zero set, in each ensemble member, $|\langle J|\varnothing\rangle |^{2}=\langle J|J\rangle\langle \varnothing|\varnothing\rangle$. Note that this holds for any state $\ket J$. This implies that in each theory, all normalized probabilities are trivial:
\be 
P(\varnothing\to J)=\frac{| \langle J|\varnothing\rangle|^{2}}{\langle J|J \rangle \langle \varnothing|\varnothing\rangle } = 1.
\label{eq:hhbadv1}
\ee
This result is a consequence of the well-known fact that, in the statistical approach, the Hilbert space of closed universes is one-dimensional in each realization of the ensemble \cite{Penington:2019npb,marolf_transcending_2020,usatyuk_closed_2024,Usatyuk:2024isz,abdalla_gravitational_2025,harlow_quantum_2025}.

\section{Inflationary Cosmology as an Explicit Example}
\label{sec:inf}
Inflation~\cite{Guth:1980zm,Linde:1981mu,Linde:1983gd,baumann_tasi_2012} is the leading explanation of the spatial flatness of the observable universe. Inflation can also explain the large scale homogeneity and isotropy of the universe. It can seed primordial density perturbations \cite{Mukhanov:1981xt,Hawking:1982cz,Starobinsky:1982ee} consistent with the observed cosmic microwave background and large scale structure \cite{Planck:2018vyg}. The Hartle-Hawking proposal is usually applied to cosmology in the context of inflationary cosmology. In this section, we will illustrate our results in this concrete setting.  

In Sec.~\ref{sec:model}, we present a simple model of slow-roll inflation. In Sec.~\ref{sec:oldinf}, we apply the no-boundary proposal to this model in the traditional (inconsistent) hybrid manner, recovering the infamous prediction of an empty universe~\cite{Vilenkin:1982de,linde1984,Vilenkin:1987kf,Maldacena:2024uhs}. We then exhibit the consistent application of the no-boundary proposal to slow-roll inflation in the conventional approach in Sec. \ref{sec:mminf}.

\subsection{Single-field Inflation in the Slow-roll Limit}
\label{sec:model}

The simplest inflationary models posit a single scalar field $\phi$ with a potential $V(\phi)\geq 0$. The Lorentzian action is 
\begin{align}\label{eqn:action}
    iI = \frac{1}{16\pi G_N} \int d^4 x \sqrt{-g} R - \frac{\gamma}{8\pi G_N} \int d^3 \sigma \sqrt{h} \ K - \int d^4 x \sqrt{-g} \left( \frac{1}{2}(\nabla \phi)^2 + V(\phi)\right)~,
\end{align}
where $K$ denotes the trace of the extrinsic curvature. In what follows, we will set $8 \pi G_N = 1$. Above,  we consider the action with a Gibbons-Hawking-York term defined for spacelike boundaries \cite{Gibbons:1976ue,York:1972sj}.  The coefficient $\gamma$ is unity for standard Dirichlet boundary conditions. For reasons discussed shortly, we will find it more convenient to consider $\gamma = 1/3$. This corresponds to so-called ``conformal boundary conditions," where $K$ is held fixed on the boundaries along with the conformal class of the induced metric, $h^{-1/3}h_{\mu \nu}$ \cite{Galante:2025emz}. 

In the minisuperspace approximation, the metric is restricted to the form
\begin{align}
    ds^2 = -dt^2 + a^2(t) d\Omega_3^2~,
\end{align}
in Lorentzian signature, where $d\Omega_3$ denotes the metric on a unit three-sphere; and the inflaton field $\phi(t)$ is assumed to depend only on time. With these restrictions, Dirichlet boundary conditions specify $(a,\phi)$. Conformal boundary conditions specify $(K,\phi)$, where $K=3\dot a/a$ is the trace of the extrinsic curvature~\cite{Bousso:1998na}.

\begin{figure}
    \centering
    \includegraphics[width=0.75\linewidth]{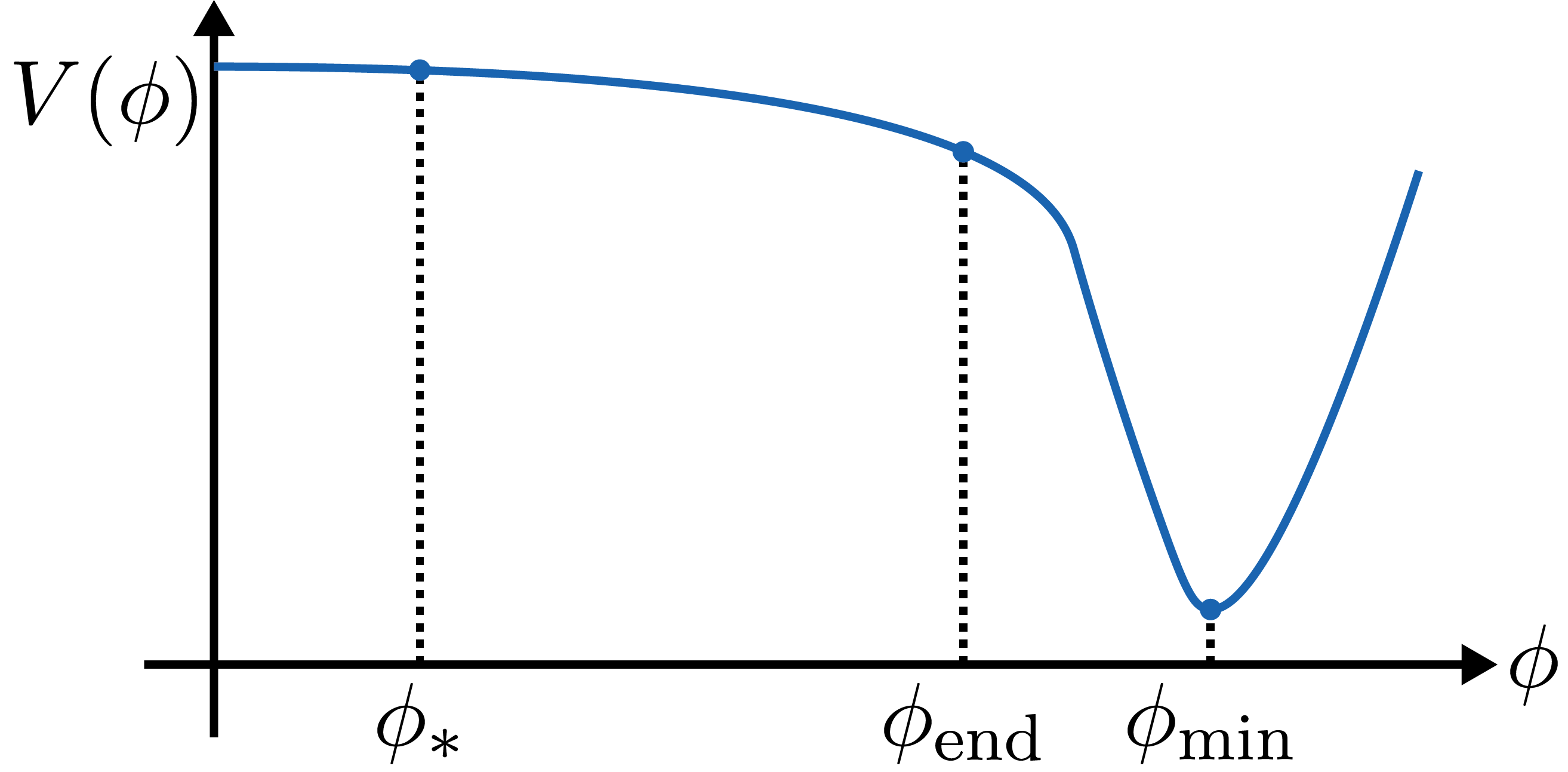}
    \caption{ During inflation, the universe undergoes exponential expansion as the inflaton slowly rolls down its potential. When $\phi\sim\phi_{\rm end}$, the potential energy of the inflaton becomes comparable to its kinetic energy, and inflation ends. The inflaton then oscillates around the minimum of the potential and decays into standard model particles (reheating phase). The value $V(\phi_{\text{min}})$ of the potential at the minimum sets the value of the cosmological constant.
    }
    \label{fig:hh_inflation}
\end{figure}

In the slow-roll approximation, one chooses a potential such that  
\begin{align}
    \epsilon = \left(\frac{dV/d\phi}{V(\phi)}\right)^2 \ll 1, \quad\quad \eta=\left(\frac{d^2V/d\phi^2}{V(\phi)}\right)\ll 1
\end{align}
over some range of $\phi$; see Fig.~\ref{fig:hh_inflation}. In what follows, we will work up to order $\epsilon$ and order $\eta^0$. In this regime, the potential energy will dominate over the kinetic energy while $\phi$ slides towards a local minimum. The universe behaves approximately as de Sitter space with a slowly varying effective cosmological constant $V(\phi)$. Near the local minimum, at some $\phi_{\rm end}$, the slow-roll approximation breaks down and $\phi$ begins to oscillate. The inflaton can then decay into radiation, connecting to standard big bang cosmology.

\subsection{Historical Application of Hartle-Hawking}
\label{sec:oldinf}

Historically, it has been assumed that the Hartle-Hawking wavefunction is proportional to $G(J)$, 
\begin{equation}
    \Psi(J)\stackrel{?}{\propto} G(J)~,\label{eq:hist}    
\end{equation}
where $J$ is an appropriate family of boundary conditions imposed at the end of inflation or during the post-inflationary era. Here we choose $J=(K,\phi_{\rm end})$, thus effectively computing a wavefunction $\Psi(K)$. The probability for different values of $K$ at the end of inflation was taken to be proportional to $|\Psi(K)|^2$. 

There exists a one-parameter family of classical Lorentzian solutions that are approximately of the form 
\begin{align}\label{eqn:clsol}
    a(t) & = H^{-1} \cosh Ht \Leftrightarrow K(t) = 3 H\tanh Ht~,\\
 \phi(t) &= \phi_* + \sqrt{\epsilon} \varphi(t)~,
\end{align}
where $H^2=V(\phi)/3$, and $\phi_*$ is an arbitrary value of $\phi$ in the slow-roll regime of $V(\phi)$. This solution also has an approximately real Euclidean section: with $\tau = \frac{\pi}{2}+ it$, 
\begin{align}\label{eqn:metric}
    a(\tau) & = H_*^{-1} \sin H_*\tau~,\\
 \phi(\tau) &\approx \phi_*~.
\end{align}
Here we neglect a small imaginary part of the Euclidean solution; see~\cite{Lyons:1992ua,Bousso:1995cc} for details. 

For every $\phi_*$ in the slow-roll range of $V(\phi)$, a solution with a single boundary $J= (K,\phi_{\rm end})$ can be obtained by joining a Euclidean contour of the form $0\leq \tau\leq \pi/2$, $t=0$ to a Lorentzian contour $0\leq t\leq t(\phi = \phi_{\rm end})$. The real part of the action of this solution is 
\begin{equation}
    \Re \,I(\phi_*) = -\frac{12\pi^2}{V(\phi_*)}~.
\end{equation}
The imaginary part will depend on details of the potential and will not be important below.

There are additional solutions such that $\dot\phi\neq 0$ at $t=0$ when $a$ is minimal. These solutions do not admit an everywhere regular Euclidean continuation. Since we are specifying only a single boundary $J=(K,\phi_{\rm end})$, these solutions do not satisfy the requirements of the no-boundary proposal and they do not contribute to $G(J)$. However, such solutions contribute a connected component to the GPI when two boundaries are specified, a case that will be discussed in Sec.~\ref{sec:mminf} below.

Physically, $\phi_*$ is the value of the inflaton at the beginning of inflation. The number of e-folds produced by inflation is a monotonic function of $|\phi_*-\phi_{\rm end}|$. Empirically, to explain the tight observational bounds on spatial curvature, we know that the number of e-folds must have exceeded approximately 60. The precise critical value depends on the details of the inflationary model. 

Another way of phrasing this empirical constraint is as the statement that at the end of inflation, when $\phi=\phi_{\rm end}$, the extrinsic curvature must have reached the value corresponding to flat slices of de Sitter, $K=3H$, up to an exponentially small correction.

Hence, with the historical approach, one finds that
\begin{align}
    \psi(K,\phi_{\rm end}) \stackrel{?}{\propto} G[K,\phi_{\rm end}] &= \exp[-I(\phi_*(K),\phi_{\rm end})]~;\\
    P(\varnothing\to (K,\phi_{\rm end})) = |\psi(K,\phi_{\rm end})|^2 &\stackrel{?}{\propto} \exp\frac{24\pi^2}{V[\phi_*(K)]}~,\label{eq:probinfold}
\end{align}
where $\phi_*(K)$ is the value of $\phi_*$ that results in the specified value of $K$ at the end of inflation when $\phi=\phi_{\rm end}$. 

The probability distribution \eqref{eq:probinfold} favors a short duration of inflation by a doubly-exponentially large factor. Qualitatively, we see this by noting that $V(\phi)$ decreases monotonically during inflation. Therefore, starting with a value of $\phi_*$ close to the end of inflation results in a larger probability. 

To see the magnitude of the problem, consider a simple model such as $V(\phi) = m^2\phi^2/2$. The mass $m$ must be of order $10^{-6}$ for the model to produce the small primordial density perturbations needed for consistency with observation. One finds~\cite{Bousso:1995cc} that $\phi(t)\approx \phi_*-\frac{m}{\sqrt{3}} t$. The number of e-folds is $N_e\approx \phi_*^2/\sqrt{3}$. Inflation ends when $\epsilon\sim 1$ at $\phi_{\rm end}\sim O(1)$. Hence
\begin{equation}
    \frac{p(K_{>60\,\rm{e-folds}})}{p(K_{O(1)\,\rm{e-folds}})}\sim 
    \frac{\exp(10^{12}/70)}{\exp(10^{12})}\sim \exp(- 10^{12}).
\end{equation}
Since the probability for the observed universe is negligibly small, this framework is ruled out.

In deference to the historical context, we have assumed that the cosmological constant vanishes after the end of inflation. If the local minimum of the potential is positive, $V(\phi_{\rm min})>0$, then the most likely universe is created with $\phi_*=\phi_{\rm min}$ \cite{Vilenkin:1987kf}. This produces an empty de Sitter spacetime with cosmological constant set by $V(\phi_{\rm min})$. Observations since the late 1990s indicate that $V(\phi_{\rm min})\sim 10^{-123}$~\cite{Planck:2018vyg}. Hence, in the historical approach to evaluating the Hartle-Hawking proposal, the observed universe would be unlikely at the level $\exp(-10^{123})$:
\begin{equation}
    \frac{p(K_{>60\,\rm{e-folds}})}{p(\text{empty~de~Sitter})}\sim 
    \frac{\exp(10^{12}/70)}{\exp(10^{123})}\sim \exp(- 10^{123}).
\end{equation}

Of course, an empty universe would be observed by no-one. In cosmology, any predicted probability distribution must be conditioned on the existence and location of observers. However, anthropic conditioning is unable to rescue a theory that assigns probabilities that grow at a doubly-exponential rate away from the observed universe. With Eq.~\eqref{eq:hist}, a universe with observers has doubly-exponentially small probability to contain anything else, such as the other planets, stars, and galaxies we observe~\cite{Dyson:2002pf,Bousso:2006xc,Page:2006hr}. Thus, the predictions extracted from the Hartle-Hawking proposal by the historical approach conflict strongly with observation.

\subsection{Consistent Evaluation}
\label{sec:mminf}

Probabilities in the statistical approach are trivially exactly one, as explained in Sec. \ref{sec:nonpert-prob}. But we also found that the probability of finding the universe in the state $\ket{J}$ is nearly one even in the conventional approach. We now verify this surprising result by explicit computation in the single-field slow-roll inflationary model of this section. 

In particular, we will exhibit the two contributions from the previous subsections that we referred to as the disconnected and connected contributions to the overlap between states of fixed extrinsic curvature, $K$, and inflaton, $\phi$. In the computation of the overlap, $\braket{K', \phi'}{K,\phi}$, we will consider $\phi$ and $\phi'$ close enough together that both are in the slow-roll region of the potential. In particular, we will take their deviation from each other to be of order the slow roll parameter, $\epsilon$, so that $\phi = \phi_* + \sqrt{\epsilon} \ \varphi$ and $\phi' = \phi_* + \sqrt{\epsilon}\  \varphi'$. Since we are constraining $\phi-\phi'$  to be close to each other, our boundary conditions are not the most general; however, they will be sufficient for illustrating our key result. 

The disconnected contribution to the overlap consists of the product of the two ``hemispheres" ending on the specified boundaries. For real Lorentzian extrinsic curvatures, these hemispheres will contain both a Euclidean portion and a Lorentzian portion of the geometry. Using the action above in Eq. \eqref{eqn:action} with $\gamma = 1/3$ and $a(t)$ as in Eq. \eqref{eqn:clsol}, a short computation gives\footnote{Recall that $\cN$ denotes a universal factor, the norm of the Hartle-Hawking state, which cancels out of probabilities. In order to display only the nontrivial contributions to amplitudes, we include appropriate factors of $\cN$ in formulas in this subsection.}
\begin{align}\label{eqn:disconnected}
    \cN^{-1} \braket{K',\varphi'}{K,\varphi}\vert_{\text{disconnected}} &\approx \inlinefig[4]{Figures/Kphi_HH.png}\times\inlinefig[4]{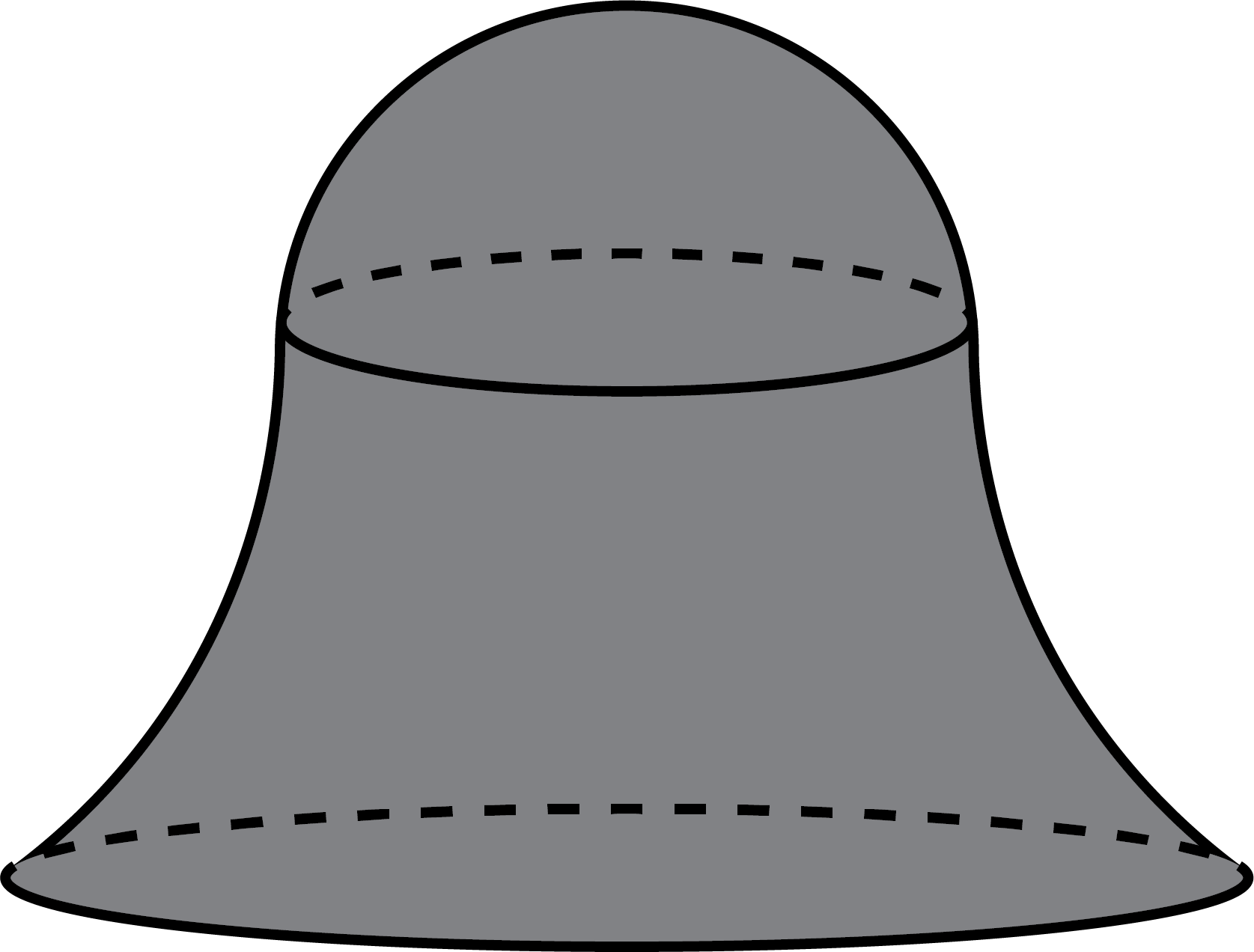} \nn\\
    &= \exp \left( \frac{24 \pi^2}{V_*}\right) \times \exp \left(i\frac{12\pi^2}{V_*} \left( \frac{\kappa}{\sqrt{1-\kappa^2}}-\frac{\kappa'}{\sqrt{1-\kappa'^2}}\right) \right)\quad, 
\end{align}
where we introduced the variable $\kappa \equiv K/\sqrt{3 V_*}$. The dependence on inflaton values $\varphi, \varphi'$ contributes phases of order $\epsilon$ to the above answer that are uninteresting. The key point for us is the exponential enhancement of the disconnected piece.

As for the connected pieces, for states of the same York time $K$, the only solution that can contribute corresponds to two slices that lie on the same side of the bounce in $a(t)$. 
In this case, as $(K, \varphi) \to (K', \varphi')$, one might wonder about ``contact" divergences associated with the proper distance between the slices going to zero. Indeed, as we show in Appendix \ref{app:inflationmodel}, if we fix $\varphi \neq \varphi'$ and bring $K \to K'$, then the contribution from this ``pancake" saddle takes the form\footnote{The inner product of states with different York times $K$ and $K'$ is also given by \eqref{eqn:osactionmain}, but the term of $O(\epsilon)$ is not divergent and hence irrelevant. As a result, the connected contribution is negligible compared to the disconnected contribution even without any smearing.} 
\begin{align}\label{eqn:osactionmain}
  &\cN^{-1}\braket{K',\varphi'}{K,\varphi}\vert_{\text{connected}} \approx \inlinefig[3]{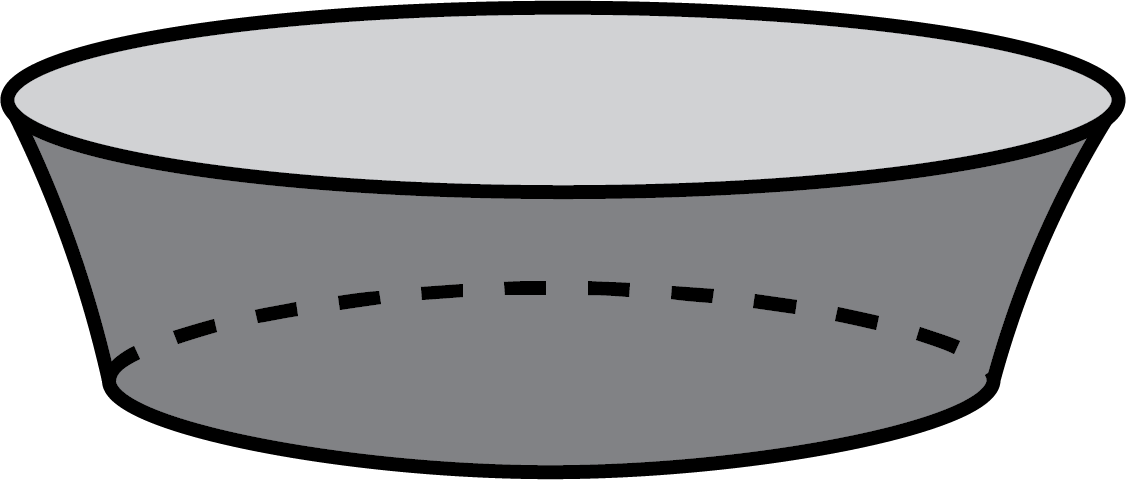} \nn\\
  &= \exp\left[\pm i \left(\frac{12\pi^2}{V_*} \left( \frac{\kappa}{\sqrt{1-\kappa^2}}-\frac{\kappa'}{\sqrt{1-\kappa'^2}}\right)+ \epsilon \pi^2 \sqrt{\frac{V_*}3} \int_{a(K)}^{a(K')} da \ a^4 \left(\frac{d \varphi}{d a}\right)^2 \right)\right] \nonumber \\
    & \approx\exp\left[\pm i\left(\frac{12\pi^2}{V_*} \left( \frac{\kappa}{\sqrt{1-\kappa^2}}-\frac{\kappa'}{\sqrt{1-\kappa'^2}}\right) + \epsilon\pi^2 \sqrt{\frac{V_*}3} \frac{a^4(K)(\varphi - \varphi')^2}{a(K)-a(K')}\right)\right]
\end{align}
The divergence that appears for $K = K'$ as $\varphi \to \varphi'$ is the sign that a delta function in $\varphi$ is emerging.\footnote{This formula should be viewed analogously to what happens in nonrelativistic quantum mechanics where we compute $\bra{x}e^{iHt}\ket{x'}$ as $t \to 0$. In the semi-classical limit, the on-shell action is just $I \sim \frac{(x-x')^2}{t}$ which is the signal that $\braket{x}{x'} \sim \delta(x-x')$. To actually see the divergence as $t \to 0$, one needs the one-loop factor as well as the on-shell action.} 
In total, then we have, as $K \to K'$
\begin{align}
    \cN^{-1}\braket{K, \varphi}{K', \varphi'} \approx  \mathcal{F}_{\text{disconn.}} \exp \left( \frac{24 \pi^2}{V_*}\right) + \mathcal{F}_{\text{pancake}} \delta(\varphi - \varphi')\quad,
    \label{eq:finaloverlap}
\end{align}
where both $\mathcal{F}_{\text{disconn.}}$ and $\mathcal{F}_{\text{pancake}}$ include phases and one-loop determinants that smoothly depend on $K, K'$ and $\varphi, \varphi'$. 

The state $\ket{K, \varphi}$ is thus non-normalizable. But of course, no physical measurements determine $\varphi$ (or $\phi$) with infinite precision. To define the transition probability from the no-boundary state, we should define a projector by smearing the possible values of the inflaton $\varphi$ while fixing $K$: 
\be 
\hat \Pi_{({K, \varphi})_\sigma} = \mathcal N^{-1} \mathcal{F}^{-1} \int_{\varphi-\sigma/2}^{\varphi+\sigma/2} d\tilde\varphi  \ket{K, \tilde \varphi}\bra{K, \tilde \varphi}\,,
\label{eq:proposed-projector}
\ee
where $\mathcal{F}$ is a normalization factor. Because we have not yet constructed a Hilbert space, this is not obviously a projection operator. We will now verify that for physically sensible choices of the top-hat smearing width $\sigma$, and with appropriate normalization $\mathcal{F}$, $\hat \Pi_{({K, \varphi})_\sigma}$ indeed has eigenvalues (approximately) $1$ and $0$ only; and we will find that the associated probability $\bra\varnothing \hat \Pi_{({K, \varphi})_\sigma} \ket\varnothing$ is nearly 1, regardless of the value of $(K,\varphi)$. 

The eigenvalue equation is 
\begin{equation}
    \hat \Pi_{({K, \varphi})_\sigma}\ket{\psi}=\lambda \ket{\psi}\,.
\end{equation}
Using Eqs.~\eqref{eq:finaloverlap} and \eqref{eq:proposed-projector}, one can verify that
\begin{equation}
\label{eq:eigenstate1}
    \ket{\psi_1} = \int_{\varphi-\sigma/2}^{\varphi+\sigma/2} d\tilde\varphi\, \ket{K,\tilde\varphi}
\end{equation}
is an eigenvector with eigenvalue  
\begin{equation}\label{eq:l1}
    \lambda_1 = \frac{\mathcal{F}_\text{pancake} + \sigma \mathcal{F}_\text{disconn.}  \exp \left( \frac{24 \pi^2}{V_*}\right)}{\mathcal{F}}~,
\end{equation}
up to the non-perturbative corrections that were neglected in \eqref{eq:finaloverlap}. 
Moreover, for any choice of complex function $f(\tilde\phi)$ that integrates to zero, 
\begin{equation}
    0=\int_{\varphi-\sigma/2}^{\varphi+\sigma/2} d\tilde\varphi\, f(\tilde\varphi)~,
\end{equation}
one finds that 
\begin{equation}
\label{eq:eigenstate2}
    \ket{\psi_f} = \int_{\varphi-\sigma/2}^{\varphi+\sigma/2} d\tilde\varphi \,f(\tilde\varphi) \ket{K,\tilde\varphi}\,,
\end{equation}
is an eigenvector with eigenvalue\footnote{Here, we are restricting ourselves to finding the eigenvectors of $ \Pi_{({K, \varphi})_\sigma}$ in the span of states $\ket{K, \varphi}$ with a single closed universe connected component. We can also construct closed universe states with multiple connected components that are orthogonal to this span and consequently also orthogonal to \eqref{eq:eigenstate1} and \eqref{eq:eigenstate2} at the level of approximation used in Eq. \eqref{eq:finaloverlap}. Therefore, such states are approximate eigenstates of $ \Pi_{({K, \varphi})_\sigma}$ with approximately zero eigenvalue and do not affect our analysis.  } 
\begin{equation}
    \lambda_0 = \frac{\mathcal{F}_\text{pancake}}{\mathcal{F}}~.
\end{equation}

Recall that $\mathcal{F}_\text{disconn.}$ and $\mathcal{F}_\text{pancake}$ smoothly depend on $(K,\varphi)$ and do not depend exponentially on $G_N^{-1}$. With any physically reasonable\footnote{To get a different answer from ours, one would have to choose  $\sigma \ll \exp \left(-24 \pi^2/(G_N^2 V_*)\right)$, where we restored factors of $G_N$ for clarity. Then $\mathcal{F} = \mathcal{F}_\text{pancake}$, so $\lambda_1 \approx \lambda_2 \approx 1$ and the projector is approximately the identity. This choice is not physically viable, as the value of the inflaton would have to be determined with exponential precision. Moreover, since  $\sigma \ll \exp \left(-24 \pi^2/V_*\right)$, contributions to the gravitational path integral that we neglected in the connected contribution in \eqref{eq:finaloverlap} would become important. For instance, this can include highly oscillating terms with a period $\exp \left(-\#/V_*\right)$ as $\varphi \to \varphi'$ which would be doubly nonperturbative.}  
choice $\sigma\ll M_p$ for the smearing width $\sigma$, the exponential term in the numerator of Eq.~\eqref{eq:l1} dominates. To obtain an approximate projection operator, we choose the overall normalization $\mathcal{F} = \sigma \mathcal{F}_\text{disconn.}  \exp \left( \frac{24 \pi^2}{V_*}\right)+ 2 \mathcal{F}_\text{pancake}$. It follows that
\begin{equation}
    \lambda_1 = 1 - O\left(\exp \left(-\frac{24 \pi^2}{V_*}\right)\right)~,~~\lambda_0 =O\left(\exp \left(-\frac{24 \pi^2}{V_*}\right)\right)~.
\end{equation}
The probability for any $(K,\varphi)$ in the no-boundary state is thus
\be
P(\varnothing \to ({K, \varphi})_\sigma) = \bra{\varnothing} \hat \pi_{({K, \varphi})_\sigma} \ket{\varnothing} = 1 - O\left(\exp \left(-\frac{24 \pi^2}{V_*}\right)\right)~, 
\label{eq:prob1last}
\ee 
confirming our claim in Sec. \ref{sec:nonpert-prob}. In particular, Eq.~\eqref{eq:prob1last} confirms that $\ket{\psi_1}\approx \ket\varnothing$; that is, any projector $\pi_{({K, \varphi})_\sigma}$ projects approximately onto the no-boundary state.

\section{Discussion}
\label{sec:discussion}

\paragraph{Assessment} Our main result is as straightforward as it is startling. In the conventional framework, the Hartle-Hawking proposal does not discriminate efficiently between different states of a closed universe. In the statistical framework, it does not discriminate at all. 

Let us recap how we arrived at this conclusion. Hartle and Hawking proposed that the universe is in the no-boundary state $\ket\varnothing$, and that the GPI can be used to compute the unnormalized amplitude $\Psi(J) \equiv \braket{J}{\varnothing}$ to observe the data $J$. Consistency demands that the norms $\braket{J}{J}$ and $\braket{\varnothing}{\varnothing}$ can also be computed from the GPI. And quantum mechanics requires us to compute these norms, in order to convert unnormalized amplitudes into normalized probabilities. In particular, the norm of $\ket J$ cannot simply be assumed independent of $J$.

Here we found that the GPI-computed norm of $\ket J$ does depend sensitively on the data $J$. As a result, the normalized probability for any cosmological data set $J$ in the no-boundary state $\ket\varnothing$ is almost 1 in the conventional interpretation of the GPI. In the statistical approach, where the GPI computes averages of products of amplitudes, we found that the probability of $J$ is \emph{exactly} 1. 

In principle, these conclusions could have been reached decades ago. In practice, our study was prompted by recent developments. Chief among them was the realization~\cite{Penington:2019npb,marolf_transcending_2020,usatyuk_closed_2024,Usatyuk:2024isz,abdalla_gravitational_2025,harlow_quantum_2025} that in the statistical approach, the GPI with closed boundaries computes averages over a collection of \emph{trivial}, one-dimensional Hilbert spaces. This general result immediately implies that all normalized inner products of closed universes are 1. Our investigation began as an effort to reconcile this broad conclusion with the apparent nontrivial probabilities found in more than four decades of studying the no-boundary proposal. In Eq.~\eqref{eq:hhbadv1} we now see that the no-boundary probabilities are in fact exactly 1 in the statistical approach.

From a historical perspective, however, it is more natural to evaluate the no-boundary proposal in the ``conventional approach,'' in which the GPI is taken to evaluate a single amplitude exactly, rather than any product of amplitudes as an average. Here was an opportunity for nontrivial probabilities to appear. And indeed, we found that probabilities need not be exactly 1. This indicates that in the conventional setting, closed universes inhabit a nontrivial Hilbert space $\mathcal{H}$. It is all the more striking, then, that the probabilities we find in Eq.~\eqref{eq:main} are exponentially close to 1. 

\paragraph{Constructing distinguishable states.} Notice that the primary problem we have encountered is not with the no-boundary proposal itself, but with the fact that classically distinct data do not define orthogonal quantum states. The state vectors $\ket J$ are not only (nearly) parallel to $\ket\varnothing$, but also to one another (and even to superpositions, so long as their number of terms does not scale exponentially with $G_N^{-1}$). This implies that the transition probability from one closed universe state to any other is also approximately $1$, which disagrees with even the most basic quantum mechanical experiments.  
This problem arises so long as the GPI is used to define inner products and to construct the Hilbert space. 

For amplitudes to distinguish between the different boundary conditions $J$ relevant in cosmology, the mapping of boundary conditions to quantum states itself would have to be modified. One possible approach is to project out the no-boundary state from every physical state. In other words, to a closed universe with data $J$, one might assign the quantum state
\begin{equation}\label{eqn:perpdef}
    \ket{J}_\perp \equiv \ket J - \frac{\ket{\varnothing}\braket{\varnothing}{J}}{\braket{\varnothing}{\varnothing}}~.
\end{equation}
The states $\ket{J_\perp}$ span the subspace $\cH_\perp$ orthogonal $\ket\varnothing$ in $\cH$.

Alternative to the above modification, we can keep the standard assignment of states $\ket J$ to symplectic data $J$, but instead change the rules of the GPI. We define
\begin{equation}
    \braket{J_1^*}{J_2} = G_{\rm conn}(J_1|J_2)~,
\end{equation}
where $G_{\rm conn}(J_1|J_2)$ is defined as $G(J_1,J_2)$ but with the additional restriction that the path integral includes only geometries $M$ that connect $J_1$ to $J_2$. (Note however that not every component of $J_1$ must be connected to some component of $J_2$.) This definition implies that
\begin{equation}\label{eq:ggg}
    G_{\rm conn}(J_1|J_2) = G(J_1,J_2) - \cN^{-1} G(J_1)G(J_2)~.
\end{equation}

    In fact, the two modifications we have described are equivalent:
\begin{align}
   \vphantom{}_{\perp}{\braket{\vphantom{J}J_{1}}{\vphantom{J}J_{2}}}\vphantom{}_{\perp} & = 
       \left(\bra{J_1} - 
       \frac{\bra{\varnothing} \braket{J_1}{\varnothing}}{\braket{\varnothing}{\varnothing}}\right)
        \left(\ket{J_2} - \frac{\ket{\varnothing}\braket{\varnothing}{J_2}}{\braket{\varnothing}{\varnothing}}\right)
        \nonumber \\ & = 
        \braket{J_1}{J_2} -  
        \frac{\braket{J_1}{\varnothing}\braket{\varnothing}{J_2}}{\braket{\varnothing}{\varnothing}}
        \nonumber \\ & = 
        G_{\rm conn}(J_1^*|J_2)~,
\end{align}
where the final equality follows from Eq.~\eqref{eq:ggg}.\footnote{This rule for computing state overlaps is reminiscent of recent proposals for how to define a closed universe Hilbert space in the presence of an observer \cite{abdalla_gravitational_2025}. In that case, connectedness was enforced by requiring the observer's worldline to connect between the bra and ket. }

With the above modification, the Hartle-Hawking proposal would still be ruled out. The states $\ket{J}_\perp$ are orthogonal to $\ket\varnothing$ by construction; and $G_{\rm conn}$ does not connect $\varnothing$ to any nonempty boundary.\footnote{The idea of defining a state orthogonal to the no boundary state was also discussed by Giddings and Strominger \cite{giddings_baby_1989}, who viewed the no boundary wavefunction as a tadpole contribution that should be removed.} Thus, it would be necessary to formulate a different proposal for the state of the universe. This may re-introduce the arbitrary choices that the no-boundary proposal was designed to evade.\footnote{For example, within the above modification, one could ``mock up'' the result \eqref{eq:old-probability-HH} traditionally (but incorrectly) ascribed to the Hartle-Hawking proposal, by proposing that the universe is in the state $\ket {J_{\rm fake}}_\perp$. Here $\ket{J_{\rm fake}}$ is defined by specifying a round $S^3$-metric, in the limit of small radius $\epsilon$. With this choice, we expect that the probability $P_{\rm mod}(J_{\rm fake} \to J)$, computed using the above modified framework, will reproduce the right hand side of Eq.~\eqref{eq:old-probability-HH}.}

\paragraph{Relationship between the conventional and statistical approach.} Let us return to the conventional approach to the GPI, with no modifications. 
We recall that states $\ket J$ are defined by suitable boundary conditions $J$ on the closed manifold $\Sigma$ (which need not be connected). The GPI defines inner products and norms. Assuming appropriate positivity conditions, this constructs a Hilbert space $\mathcal{H}$. However, the existence of nontrivial null states obscures the structure of $\mathcal{H}$. In particular, our main result \eqref{eq:main} implies that states defined by distinct data $J\neq J'$ need not be orthogonal; indeed, even states with different numbers of connected components will not be orthogonal.

In order to go further and define an orthonormal basis of $\mathcal{H}$, let the operator $\hat Z[J]$ have the following action on the state $\ket {J'}$~\cite{marolf_transcending_2020}: 
\begin{align}
    \hat Z[J]\ket {J'} = \ket {J,J'}.
\end{align}
Since $\mathcal{H}$ is spanned by the states $\ket {J'}$, this defines the action of $\hat Z[J]$ on any state. In particular, \begin{equation}
    \hat Z[J] \ket\varnothing = \ket J~.
\end{equation}
We also define the adjoint operator
\begin{align}
    \hat{Z}[J]^\dagger = \hat{Z}[J^*]~.
\end{align}
Since $G(J_1, \cdots, J_n)$ is invariant under permutations of its arguments,
\begin{align}
    [\hat{Z}(J), \hat{Z}(J')]=0
\end{align}
for any pair of boundary conditions $J$ and $J'$. With the choice $J'\to J^*$, this implies that the operator $Z[J]$ is normal, for all $J$. Combining these results, it follows that the operators $Z[J]$ are simultaneously diagonalizable. They define an orthonormal basis $\{\ket\alpha \}$ of $\mathcal{H}$, with (complex) eigenvalues $Z_\alpha[J]$ \cite{marolf_transcending_2020}:
\begin{align}
    \hat{Z}[J] \ket{\alpha} = Z_\alpha[J] \ket{\alpha}~.
\end{align}
The basis states $\ket \alpha$ are fully determined (up to a phase) by the collection of operators $\hat Z[J]$. If the spectrum of the $\hat Z$ operators is continuous, the basis will be Dirac-orthonormal; we leave this implicit.

The construction of the states $\ket\alpha$ allows us to recover the statistical approach from the conventional approach. For any bra and ket, an ensemble of amplitudes indexed by $\alpha$ can be obtained by inserting a projector $\hat \pi_\alpha = \ketbra{\alpha}$ into their inner product and assigning the probability $P(\alpha) = |\braket{\alpha}{\varnothing}|^2$. The results in the statistical approach arise from taking averages of products of projected inner products in this ensemble. For example, for one inner product, we have
\begin{align}
  \overline{\langle J_1|J_2\rangle} = \int d\alpha \mel{J_1}{\hat{\pi}^{\rm CU}_\alpha}{J_2} = \int d \alpha  \ P(\alpha) \frac{\langle J_1| \hat \pi_\alpha^\text{CU}|J_2\rangle}{ \bra{\varnothing} \hat \pi_\alpha^\text{CU} \ket{\varnothing}} =   \frac{1}{\mathcal N} \braket{J_1}{J_2},
\end{align}
while for a product of two inner products, we have
\be 
\overline{\langle J_1|J_2\rangle \langle J_3|J_4\rangle} &= \int D \alpha~P(\alpha) \frac{\langle J_1| \hat \pi_\alpha |J_2\rangle}{\langle \varnothing| \hat \pi_\alpha|\varnothing\rangle} \frac{\langle J_3| \hat \pi_\alpha|J_4\rangle}{\langle \varnothing| \hat \pi_\alpha|\varnothing\rangle} = \frac{1}{\mathcal N}\braket{J_1, J_3}{J_2,J_4}\,.
\ee
We discuss this relationship in detail in Appendix \ref{sec:alphastates}.

\paragraph{Not all observables are diagonal in the $\alpha$ basis.}
In the framework described above, we can define the projector $\hat \pi_{J \perp} = \ket{J} \vphantom{}_\perp \vphantom{}_\perp \bra{J}$ whose expectation value describes the transition probability to a universe with boundary conditions $J$. Summing over such projectors with different $J$ yields observables that probe the different properties of the closed universe (e.g., its size, the state of its fields, etc.). Such operators are not diagonal in the $\alpha$ basis. This can be seen from
\be 
\bra{J_1}\left[\hat Z[ J_3], \hat\pi_{J_{4\perp}} \right] \ket{J_2} = \braket{J_1, J_3^*}{ J_{4}}\vphantom{}_\perp \vphantom{}_\perp \braket{J_{4}}{J_2} -  \braket{J_1}{J_{4}} \vphantom{}_\perp \vphantom{}_\perp\braket{J_{4}}{J_3, J_2}\neq 0 \,,
\label{eq:non-diag-alpha}
\ee
for generic $J_1$ and $J_2$ since the GPI contributions to the first and second terms are different.

In the context of AdS/CFT, boundary observables do not mix the $\alpha$-sectors constructed by Marolf and Maxfield~\cite{marolf_transcending_2020}. Such observables are associated with CFT operators. As with $\pi_{J \perp} $, one can also write down operators that change the value of $\alpha$.  An important physical question is whether observations by bulk observers are ever associated with such non-diagonal operators. Eq.~\eqref{eq:non-diag-alpha} shows that, within the approach pursued in this discussion, operators relevant for observers inside a gravitating spacetime are not diagonal in the $\alpha$ basis.

\section*{Acknowledgments}

 We thank Daniel Harlow, Tom Hartman, Henry Lin, Andrei Linde, Juan Maldacena, Don Marolf, Henry Maxfield, Jacob McNamara, Don Page, Douglas Stanford, Zixia Wei, Wayne Weng, and Zhenbin Yang for discussions and correspondence. The work of AIA, SA, RB, and LVI is supported in part by the Leinweber Institute for Theoretical Physics at UC Berkeley; by the Department of Energy, Office of Science, Office of High Energy Physics through awards DE-SC0025522 and DE-SC0025293; and by the Department of Energy through QuantISED award DE-SC0019380. AIA is supported by the National Science Foundation through Grant No.\ DGE-2146752. AL is supported by the John Templeton Foundation Award 41001491-013, and by the Department of Energy under QuantISED Award DE-SC0025937.

\subsection*{}

\appendix

\section{$\ket{\alpha}$-states and the Relation Between the Conventional and Statistical Approaches to the Gravitational Path Integral}
\label{sec:alphastates}

In this appendix, we discuss the relationship between the conventional and statistical approach to the GPI, using the Marolf-Maxfield construction of a closed universe Hilbert space~\cite{marolf_transcending_2020}.  We will begin by expanding on our discussion, in Sec.~\ref{sec:discussion}, of the $\alpha$ states in a theory of closed universes. We will then consider a larger Hilbert space which includes open universes
and construct a similar $\alpha$-basis. In all cases, the statistical interpretation is recovered from the conventional approach by averaging over appropriate projections onto $\alpha$-sectors inserted in inner products. We will contrast the case of open universes, encountered in AdS/CFT, with the case of closed universes.

\subsubsection*{Closed Universes}

Following \cite{marolf_transcending_2020}, to define $\alpha$-states we need to introduce the operators $\hat{Z}[J_i]$ that act on states in the Hilbert space of closed universes, $\mathcal H^\varnothing$, and create an additional closed slice with boundary conditions $J_i$,
\begin{equation}
    \ket{J_1 \dots J_n} = \hat{Z}[J_n]\ket{J_1 \dots J_{n-1}} =  \hat{Z}[J_n] \dots  \hat{Z}[J_1] \HH.
 \end{equation}
Since the inner products between such states described in Sec. \ref{sec:inner-prod-review} are independent of the ordering among the different boundary conditions on the closed connected components of $J$, all such operators commute,
\begin{equation}
\label{eq:commutators-CU-Zs}
    [\hat{Z}[J],\hat{Z}[J']] = 0\,,
\end{equation}
regardless of the boundary condition associated with the closed slice that they create. Since,  $\langle J_1 \dots J_n  |\tilde J_1\dots  \tilde J_n \rangle = \langle J_1 \dots J_{n-1}  |J_n^* \tilde J_1\dots  \tilde J_n\rangle $, the Hermitian adjoint of such operators is defined as
\begin{equation}
    \hat{Z}^\dag[J] = \hat{Z}[J^*]\,.
\end{equation}
Using \eqref{eq:commutators-CU-Zs}, $ [\hat Z[J],\hat Z^\dag[J']] = 0$ follows. Since all operators $\hat{Z}[J]$ commute and are normal, we can find a common complete orthogonal basis of eigenstates $\ket{\alpha}$ with eigenvalues $Z_{\alpha}[J]$,
\be
    &\hat{Z}[J]\ket{\alpha}=Z_\alpha[J]\ket{\alpha} \qquad \forall J\,,
\ee
with $\mathcal H^\varnothing  = \text{span}\left\{\ket{\alpha}\right\}$. For concreteness, we will assume that the eigenstates $\ket{\alpha}$ form a continuous set of states and are delta-function normalized, such that $\braket{\alpha}{\alpha'}=  \delta(\alpha-\alpha')$ such that $\int d\alpha \,\delta(\alpha-\alpha')\,f_\alpha = f_{\alpha'}$ is valid with a flat measure. This allows us to write
\be
     \frac{\braket{\tilde J_1 \dots \tilde J_{\tilde n}}{J_1 \dots J_n}}{\braket{\varnothing}{\varnothing}} = \int d \alpha P(\alpha) Z_\alpha[\tilde J_1^*] \dots Z_\alpha[\tilde J_{\tilde n}^* ]Z_\alpha[J_1] \dots Z_\alpha[J_n]\,,
     \label{eq:inner-product-avg-over-alpha}
\ee
where \be 
P(\alpha) = \frac{|\braket{\alpha}{\varnothing}|^2}{\braket{\varnothing}{\varnothing}}, 
\ee
Since $P(\alpha) >0$ and $\int d\alpha P(\alpha) = 1$, we can thus view $P(\alpha)$ as a probability distributions. Consequently, the inner products in Eq. \eqref{eq:inner-product-avg-over-alpha} are given by the average of the product of $Z_\alpha[\dots]$ 
within this probability distribution. As we shall explain shortly, this rewriting of the inner product will allow us to establish a concrete connection between the two approaches discussed in Sec. \ref{sec:inner-prod-review}.

\subsubsection*{Open Universes}

To understand the usefulness of $\alpha$-states, it is also instructive to define states on open universe slices for which we will once again follow \cite{marolf_transcending_2020}. These are specified by a set of boundary conditions $\cJ$ a manifold $\Sigma $ that itself has a boundary partial $\partial\Sigma$. As in the case of closed universes, if $\Sigma$ consists of multiple connected components, we shall denote the boundary conditions on each connected component by $\cJ_i$.

To define the boundary condition for the GPI associated with an overlap 
\begin{equation}
 \braket{\cJ}{\tilde\cJ} \equiv \langle \mathcal{J}_1,...,\mathcal{J}_n|\tilde{\mathcal{J}}_1,...,\tilde{\mathcal{J}}_n\rangle, 
\end{equation} 
one must first glue the open boundaries in pairs to form closed boundaries. Two open boundaries $\Sigma_i$, $\tilde{\Sigma}_i$ can only be glued to each other if their boundary conditions on $\partial\Sigma_i$, $\partial\tilde{\Sigma}_i$ agree. In other words, we want $\mathcal{J}_i^*$ and $\tilde{\mathcal{J}}_i$ to satisfy $\partial\mathcal{J}_i^*=\partial\tilde{\mathcal{J}}_i$, where $\partial \cJ$ is the boundary condition on $\partial \Sigma$ induced for $\cJ$. The simplest example is when the induced metric on  $\partial\Sigma_i$ and $\partial\tilde{\Sigma}_i$ agree. If two open universes do not share the same boundary conditions, then an inner product cannot be defined -- this is the reason why we included the superscript $\partial \Sigma$ when defining the state. The closed boundary condition, which we denote by $J_i$, resulting from gluing together two open boundaries, is given by\footnote{Here, we restrict ourselves to the case where we solely consider gluings of $\partial \mathcal{J}_i^*$ to $\partial \tilde{\mathcal{J}}_i$. There could be multiple boundaries among $\{\partial \mathcal{J}_k^*, \partial \tilde{\mathcal{J}}_k\}$  of the open universes slices that have the same boundary conditions and, consequently, can be glued in different ways. In such a case, each pairing among $\{\partial \mathcal{J}_k^*\}$ and $\{\partial \tilde{\mathcal{J}}_k\}$ would define a different inner product, and we will solely be focusing on one such pairing.}
\be
    J_i=  \cJ_i^*\cup \tilde \cJ_i \equiv  \cJ_i^*\sqcup \tilde \cJ_i/(\partial \cJ_i^*\sim\partial \tilde\cJ_i)\,.
\ee
After gluing all pairs of open boundaries, the GPI computes the overlap 
\be
 \langle \mathcal{J}_1,...,\mathcal{J}_n|\tilde{\mathcal{J}}_1,...,\tilde{\mathcal{J}}_n\rangle=\langle \varnothing|\mathcal{J}_1^*\cup \tilde{\mathcal{J}}_1...\mathcal{J}_n^*\cup \tilde{\mathcal{J}}_n\rangle=G(\mathcal{J}_1^*\cup \tilde{\mathcal{J}}_1,...,\mathcal{J}_n^*\cup \tilde{\mathcal{J}}_n)\,.
\ee
Since inner products between states with $\partial \tilde \cJ^* \neq \partial \cJ$ are ill-defined, the underlying Hilbert space is labeled by the ordered list of boundaries,
\be \ket{{ \cJ_1,\, \dots,\,  \cJ_{ n}}} \in  \mathcal H^{\partial \cJ}  \equiv H^{(\partial \cJ_1, \dots, \partial \cJ_k)}\,.
\label{eq:Hilbert-space-state-OU}
\ee
Note that closed universes are simply a special case of the equations above when choosing $\partial \cJ = \partial \tilde \cJ = \varnothing$. In the above notation, the closed universe Hilbert space is denoted by $\mathcal H^{CU} = \mathcal H^\varnothing$.
Notice that, unlike for closed universe states, the ordering of the $\mathcal{J}_i$ in the definition of open universe states is crucial because it determines the pairings between open boundaries in the bra and the ket when computing inner products: different orderings correspond to distinct states that might not even belong to the same Hilbert space (e.g., if $\partial J_1 \neq \partial J_2$, then $\mathcal H^{(\partial \cJ_1, \partial \cJ_2, \dots, \partial \cJ_k)} \neq \mathcal H^{(\partial \cJ_2, \partial \cJ_1,\partial\cJ_3, \dots, \partial \cJ_k)} $). 

It should now be clear that the same ambiguity in interpretation we observed for closed universe states is also present for open universe states. The difference between the two approaches can be summarized as follows 
\begin{equation}
\label{eq:MM-OU-inner-prod}
    G(J) = \begin{cases} \overline{\langle \mathcal{J}_1, \dots, \mathcal{J}_n|\tilde{\mathcal{J}}_1, \dots,\tilde{\mathcal{J}}_n\rangle} =\overline{\langle \mathcal{J}_1|\tilde{\mathcal{J}}_1\rangle \dots \langle \mathcal{J}_n|\tilde{\mathcal{J}}_n\rangle},\\ 
    \langle \mathcal{J}_1, \dots, \mathcal{J}_n|\tilde{\mathcal{J}}_1, \dots,\tilde{\mathcal{J}}_n\rangle \neq \langle \mathcal{J}_1|\tilde{\mathcal{J}}_1\rangle \dots \langle \mathcal{J}_n|\tilde{\mathcal{J}}_n\rangle,
    \end{cases}
\end{equation}
where $J = (\cJ_1 \cup \tilde \cJ_1, \dots, \cJ_n \cup \tilde \cJ_n)$. 

We can also define states with both open and closed universe components, for instance $\ket{{J_1 \dots J_n;  \cJ_1,\, \dots,\,  \cJ_{ n}}} \in \mathcal H^{(\partial \cJ_1, \dots, \partial \cJ_k)} $. In this case, because the boundary conditions on the boundary of $\Sigma$ are the same as in \eqref{eq:Hilbert-space-state-OU}, such states live within the same Hilbert space as \eqref{eq:Hilbert-space-state-OU}.
The inner product in such a case is given by, 
\be 
\label{eq:MM-OU-CU-inner-prod}
  &G\left(\tilde  \cJ_1^* \cup \cJ_1, \dots, \tilde \cJ_n^* \cup \cJ_n,\, \tilde J_1^*,\,\dots,\,\tilde J_{\tilde n}^*, \,J_1, \dots,\, J_n\right)  \nn = \\ &\qquad =\begin{cases}
  \overline{\braket{\tilde J_1 \dots \tilde J_{\tilde n}; \tilde \cJ_1,\, \dots,\, \tilde \cJ_{ n}}{J_1 \dots J_n; \tilde \cJ_1,\, \dots,\, \tilde \cJ_{ n}}},\\ \braket{\tilde J_1 \dots \tilde J_{\tilde n}; \tilde \cJ_1,\, \dots,\, \tilde \cJ_{ n}}{J_1 \dots J_n; \tilde \cJ_1,\, \dots,\, \tilde \cJ_{ n}}.
  \end{cases}
\ee

In the conventional approach, $\alpha$-states play an equally important role for the Hilbert space of states with open universes. 
As with closed universes, we can define an operator $\hat \psi[\cJ]$ that adds an additional open universe with boundary condition $\cJ$ to any states. Since this operator changes the number of open universe components, $\hat \psi[\cJ]: \mathcal H^{(\partial \cJ_1, \dots, \partial \cJ_k)} \rightarrow \mathcal H^{(\partial \cJ, \partial \cJ_1, \dots, \partial \cJ_{k})}$. Thus, starting from the Hartle-Hawking state, we can construct a state in  $\cH^{(\partial \cJ_1, \dots, \partial \cJ_k)} $, by acting with such operators multiple times, we get 
\be 
\ket{\cJ_1, \, \dots,\, \cJ_n} =  \hat \psi[\cJ_1]\dots  \hat \psi[\cJ_n]\HH\,.
\ee
Note that the ordering of operators on the right-hand-side is important in order to define a state in  $H^{(\partial \cJ_1, \dots, \partial \cJ_n)} $.
One can also similarly define $\hat\psi[\tilde \cJ]^\dagger: \cH^{(\partial \cJ, \partial \cJ_1, \dots, \partial \cJ_{k})} \to \cH^{( \partial \cJ_1, \dots, \partial \cJ_{k})}$, where $\partial \tilde\cJ^* = \partial \cJ$, as the operator that, when acting on a state with an open universe slice with boundary $\partial \tilde \cJ^*$, glues to it an open universe with boundary conditions $\cJ$ (with $\partial \cJ^* = \partial \cJ$), 
\be 
\hat\psi[\tilde \cJ]^\dagger \ket{\cJ,  \cJ_1,\, \dots,\,  \cJ_{ n}} = \ket{ \tilde\cJ^* \cup \cJ; \,\cJ_1, \, \dots,\,\cJ_n}\,.
\ee
Using these operations sequentially, we find that $\bra{\varnothing} \hat \psi^\dagger[\tilde \cJ_n] \dots  \hat \psi^\dagger[\tilde \cJ_1] \hat{\psi}[ \cJ_1] \dots  \hat \psi[ \cJ_n] \ket{\varnothing}$ reproduces the inner product in  \eqref{eq:MM-OU-inner-prod}.

Just like in the closed universe Hilbert space, in $\cH^{( \partial \cJ_1, \dots, \partial \cJ_{k})}$ we can still simultaneously diagonalize all operators $\hat Z[J]$, 
\be  
\hat Z[J] \ket{\alpha; \cJ_1, \, \dots,\,\cJ_n}=  Z_\alpha[J]  \ket{\alpha; \, \cJ_1, \, \dots,\, \cJ_n}\,, \qquad \forall J\,.
\ee
Using the rules above, the inner product between two such states is 
\be 
\braket{\tilde\alpha; \, \tilde \cJ_1, \, \dots,\, \tilde \cJ_n}{\alpha; \, \cJ_1, \, \dots,\, \cJ_n} = \delta(\alpha-\tilde \alpha) Z_\alpha[\tilde \cJ_1^* \cup \cJ_1] \dots, \, Z_\alpha[\tilde \cJ_n^* \cup \cJ_n]\,.
\ee
This makes it clear that the Hilbert space splits as, 
\be 
\label{eq:OU-Hilbert-space-decomposition}
\ket{\cJ_1, \, \dots,\, \cJ_n} \in \cH^{( \partial \cJ_1, \dots, \partial \cJ_{k})} =  \oplus_\alpha (\cH_\alpha^{\partial \cJ_1 } \otimes \dots \otimes \cH_\alpha^{\partial \cJ_n })\,.
\ee
where $\cH_\alpha^{\partial \cJ} $ denotes a Hilbert space that is labeled by the boundary condition on $\partial \Sigma$, $\partial \cJ$. 
This is consistent with a probabilistic  interpretation of $\alpha$-states we discussed above: each $\cH_\alpha^{\partial \cJ_1 } \otimes \dots \otimes \cH_\alpha^{\partial \cJ_n }$ is the Hilbert space on $(\partial \cJ_1, \, \dots, \, \partial \cJ_n)$ of individual theories labeled by $\alpha$. Then, just as for closed universes
\be
     \frac{\langle \tilde{\mathcal{J}}_1,...,\tilde{\mathcal{J}}_n|{\mathcal{J}}_1,...,{\mathcal{J}}_n\rangle}{\braket{\varnothing}{\varnothing}} = \int d \alpha P(\alpha) Z_\alpha[\tilde \cJ_1^* \cup \cJ_1] \dots, \, Z_\alpha[\tilde \cJ_n^* \cup \cJ_n]\,,
     \label{eq:inner-product-avg-over-alpha-open} 
     \ee
     which thus computes the inner products in the individual theories labeled by $\alpha$, and then averaging over the theories with an appropriate probability distribution $P(\alpha)$. In the context of AdS/CFT, $\alpha$ labels each member of the ensemble in a putative ensemble of CFTs, and $\cH_\alpha^{\partial \cJ}$ is the Hilbert space of CFT$_\alpha$ on the spatial slice $\partial \cJ$. We will now use this fact to explain the relation between the two approaches. 

     \subsubsection*{Relation Between the two Approaches}

     Given the structure of the Hilbert space in \eqref{eq:OU-Hilbert-space-decomposition}, it will be useful to define a projector onto the open or closed universe Hilbert space labeled by $\alpha$, such that 
\begin{align}
    \hat{\pi}_{\alpha}:\cH^{( \partial \cJ_1, \dots, \partial \cJ_{k})} \to  \cH_\alpha^{\partial \cJ_1 } \otimes \dots \otimes \cH_\alpha^{\partial \cJ_n }\,.
\end{align}

Within the conventional approach, any open-universe calculation done in the statistical approach can then be understood as computing expected values for open-universe measurements before a specific $\alpha$ state has been measured. This corresponds to computing the average over $\alpha$ of open-universe measurements performed in a fixed $\alpha$ sector. 
For example, using the projector $\hat{\pi}_{\alpha}$, we can write the average inner product between two states as
\be \label{eqn:OUIP}
\overline{\langle \cJ_1|\cJ_2\rangle} = \int D \alpha  \ P(\alpha)  \frac{\langle \cJ_1| \hat \pi_\alpha|\cJ_2 \rangle}{\langle \varnothing| \hat \pi_\alpha|\varnothing\rangle} = \frac{1}{\mathcal N}{\langle \cJ_1|\cJ_2\rangle}.
\ee
Fluctuations of this inner product over the set of possible $\alpha$ states can also be computed analogously by the expression
\be 
\overline{\langle \cJ_1|\cJ_2\rangle \langle \cJ_3|\cJ_4\rangle} &= \int D \alpha~P(\alpha) \frac{\langle \cJ_1| \hat \pi_\alpha |\cJ_2\rangle}{\langle \varnothing| \hat \pi_\alpha|\varnothing\rangle} \frac{\langle \cJ_3| \hat \pi_\alpha|\cJ_4\rangle}{\langle \varnothing| \hat \pi_\alpha|\varnothing\rangle} \nn\\  &= \frac{1}{\mathcal N}\braket{\cJ_1, \cJ_3}{\cJ_2,\cJ_4}\,,
\ee
where in the second equality we emphasized that, within the conventional approach, the same GPI calculation can be interpreted as computing a single inner product between two-universe states.
Extending these calculations to higher moments, it should be clear that any calculation performed within the statistical approach is expressible within the conventional formalism as an average over the ensemble of $\alpha$ sectors. Using these statistics over $\alpha$ sectors, one can infer properties of the inner product and of the Hilbert space in a single (typical) $\alpha$ sector.

The equations above also apply to closed universes by choosing $\cJ =  J$ with $\partial \cJ = \varnothing$.  In this notation, the statistical average of the inner product between two closed-universe states is 
\be 
\label{eqn:CUIP}
\frac{1}{\mathcal N}\overline{\langle J_1|J_2\rangle} = \int D \alpha  \ P(\alpha) \frac{\langle J_1| \hat \pi_\alpha|J_2\rangle}{ \bra{\varnothing} \hat \pi_\alpha \ket{\varnothing}}\ = \braket{J_1}{J_2} .
\ee
where it should again be clear how to extend this to higher moments of the closed universe inner product. From this relation between the inner product as computed in the two approaches, it is obvious why we get that the inferred Hilbert space of closed universes in the statistical approach is one-dimensional, whereas in the case of open universes -- in computations done in the context of AdS/CFT -- the inferred Hilbert space is nontrivial. For closed universes, the projector $\hat\pi_\alpha = \ket{\alpha}\bra{\alpha}$ is a projector of rank one, and so there is only one closed universe state in each $\alpha$-sector. In contrast, for open universes, the projector has a rank equal to the dimension of $\mathcal{H}^{\partial \cJ}_{\alpha}$, which can be nontrivial.

\section{Connected and Disconnected Saddles in a Model of Inflation}\label{app:inflationmodel}

In the main text, we discussed states specified by fixing the trace of the extrinsic curvature, the conformal class of the metric, and the inflaton. These are referred to as conformal boundary conditions. 
In this appendix, we first discuss states specified by fixing the inflaton $\varphi$ and the metric (namely $a$). This choice corresponds to Dirichlet boundary conditions. After deriving the on-shell action for Dirichlet boundary conditions, we will explain how to obtain the on-shell action for the conformal boundary conditions used in the main text and derive Eq. \eqref{eqn:osactionmain}.

We continue discussing a simple model of slow-roll inflation. In particular, we follow closely the discussion in \cite{Maldacena:2024uhs}. As for states specified by conformal boundary conditions, we will discuss connected contributions to the gravitational path integral that can contribute to the inner product. We will show that these contributions are exponentially suppressed relative to the no-boundary saddle. As in the main text, we consider the Lorentzian action in 4 space-time dimensions
\begin{align}
iI =  \int_M d^4 x \sqrt{-g}\, \left( \frac{R}{16\pi G_N} - \frac{1}2 (\nabla \phi)^2 - V(\phi)\right) - \frac{\gamma }{8\pi G_N} \int_{\partial M} K \quad,
\end{align}
where here we will set $\gamma = 1$ to fix $a$ at given spatial slices (Dirichlet boundary conditions \cite{Gibbons:1976ue,York:1972sj}), or $\gamma=1/3$ to fix the extrinsic curvature $K$ and the conformal class of the induced metric (conformal boundary conditions \cite{Galante:2025emz}). Let us first carry out the analysis for $\gamma=1$, and then we will comment on the difference when choosing $\gamma=1/3$ as in the main text. For simplicity, we set $8 \pi G_N = 1$ in what follows. The Friedmann equation and the equation of motion for $\phi$ are then
\begin{align}\label{eqn:eom}
&3 H^2 = -\frac{3}{a^2}+ \frac{1}2 \dot{\phi}^2 +V(\phi),\ \nonumber \\
& \ddot{\phi}+ 3H\dot{\phi}+ V'(\phi) = 0,\ 
\end{align}
where $H^2 = \frac{\dot{a}^2}{a^2}$. Now we analyze these equations in the slow-roll approximation where we assume  $\epsilon = \left( \frac{V_*'}{V_*}\right)^2 \ll 1$ and $\eta = \left(\frac{V_*''}{V_*}\right)\ll 1$. In the strict zero-roll limit, the potential is constant at some value $V(\phi) = V_* \equiv V(\phi_*)$ and the geometry is just given by 
\begin{align}\label{eqn:asol}
    a(t) = \frac{1}{H_*} \cosh \left( H_* t\right) \quad H_* = \sqrt{V(\phi_*)/3}.
\end{align}
We can now expand about this background solution to order $O(\sqrt{\epsilon})$ and order $O(\eta^0)$. Defining a new time $\tau = H_*t$ and expanding the scalar (inflaton) as $\phi = \phi_* + \sqrt{\epsilon} \varphi$, Eq. \eqref{eqn:eom} implies 
\begin{align}
    \partial_{\tau}^2 \varphi + 3 \tanh \tau \partial_{\tau}\varphi + 3 =0\quad.
\end{align}
The general solution to this equation is
\begin{align}\label{eqn:varphigensol}
    \varphi(\tau) = c_1\arctan \left( \tanh(\tau/2)\right) +c_2 - \log \cosh \tau+\frac{1}{\cosh^2 \tau} + \frac{c_1}2 \frac{\sinh\tau}{\cosh^2 \tau}.
\end{align}
Note that at large $\tau$ this function rapidly becomes linear in $\tau$ with a constant shift set by $\frac{\pi c_1}4+c_2+\log(2)$. 

Now we want to understand the overlap between states labeled by field values 
\begin{align}
    (a, \varphi) = (a_0, \varphi_0)\ \ \text{and}\ \ (a,\varphi) = (a_0', \varphi_0')\quad. 
\end{align}To leading order in the slow-roll limit, all the solutions with these boundary conditions will have a geometry given by (an analytic continuation of) the solution \eqref{eqn:asol}. The on-shell action for such geometries is then just the standard pure de Sitter action with a cosmological constant set by $V(\phi_*)$, with contributions from the nontrivial $\varphi$ dependence that are subleading in the $\epsilon$ expansion. As discussed in the main text, there will be four disconnected saddle points and four connected ones. These four saddle points each correspond to the four different choices of time coordinate $t$ for the location of the boundary slices at $a = a_0$ and $a_0'$. To compute the on-shell action of these saddles, we write the action in terms of $a$ and $\phi$ as 
\begin{align}\label{eqn:D7}
    iI &= \frac{V_{S^3}}{2} \int d t a^3(t) \left(\frac{6}{a^2} + 6 \frac{d^2}{dt^2} \log a + 12 \left(\frac{d}{dt} \log a\right)^2\right) + I_{\text{matter}} + I_{GHY}\nonumber \\
    & = \frac{V_{S^3}}{2} \int d t \ \left(6a - 6 \dot{a}^2 a \right) + I_{\text{matter}} 
\end{align}
where we integrated by parts in the second line which eliminated the boundary term. 

\subsection*{Disconnected Contributions}
For the disconnected contributions, the time contour for each connected component of the saddles will start/end at $\tau_{\text{cap}} = \pm \frac{i \pi}2 $ where $a(\tau_{\text{cap}}) = 0$. The sign choice for $\tau_{\text{cap}}$ depends on whether the contour is for the bra or ket in the overlap $\braket{a_0, \varphi_0}{a_0', \varphi_0'}$. One sums over all such contours and the final answer is
\begin{align}\label{eqn:D8}
    iI &= \left. \frac{6 \sqrt{a^2 H_*^2-1} + \frac{1}{\left(H_*a+\sqrt{a^2 H_*^2-1} \right)^3}+\left(H_*a+\sqrt{a^2 H_*^2-1}\right)^3}{4H_*^2}\right \vert_{a=0}^{a=a_0} + (a_0 \leftrightarrow a_0') \nonumber \\
    &\approx -i \frac{24 \pi^2}{V_*}+ 4\pi^2 \sqrt{\frac{V_*}3}\left(\mp_1 a_0^3 \mp_2 a_0'^3\right) +  6\pi^2 \sqrt{\frac{3}{V_*}}\left(\pm_1 a_0 \pm_2 a_0'\right)  + \mathcal{O}\left(\frac{1}{a_0}, \frac{1}{a_0'}\right), 
\end{align}
where in the second line we took the large $a_0$, $a_0'$ limits and used $H_* = \sqrt{V_*/3}$. The different signs represent the four different saddles, where signs with the same subscript need to be chosen consistently. The point is that this answer is pure imaginary and of order $1/V_*$, except for the oscillatory phases. There will be corrections to this answer at higher orders in $\epsilon$ that will not change the fact that this saddle contributes an exponential enhancement to the path integral with these boundary conditions. Adding the contribution of these saddles to the path integral gives the answer \eqref{eqn:disconnected} quoted in the main text. 

\subsection*{Connected Contributions}
For the connected contributions, again there are four saddles corresponding to the four solutions to $a(\tau_0) = a_0$ and $a(\tau_0') = a_0'$. As before, we will analyze when $a, a'$ is large and real so that $|\tau|$ for the slices is large and $\tau$ is mostly real and we have $(a, a') \approx \sqrt{\frac{3}{4V_*}} (e^{|\tau_0|}, e^{|\tau_0'|})$. To find the solution for $\varphi$, we just have to solve for the coefficients $c_1$ and $c_2$ in \eqref{eqn:varphigensol} by demanding
\begin{align}\label{eqn:phibc}
    \varphi(\tau_0) =\varphi_0,\ \  \varphi(\tau_0') = \varphi_0'\quad .
\end{align}
Since this is a linear system of two equations with two unknowns and real coefficients, there will be a real solution for $\varphi(\tau)$. At large $|\tau_{a,a'}|$, the form of $\varphi$ simplifies to
\begin{align}\label{eqn:philargetau}
    \varphi(\tau) = \text{sign}(\tau) \frac{c_1 \pi}4 + c_2 - |\tau|+ \log 2 +3e^{-2|\tau|} -\frac{8}{3}c_1\text{sign}(\tau)e^{-3|\tau|}+ \mathcal{O}(e^{-4|\tau|}).
\end{align}

If the signs of $\tau_0$ and $\tau_0'$ are different so that the two slices of volume $a$ and $a'$ are on opposite sides of the bounce in \eqref{eqn:asol}, then we can solve \eqref{eqn:phibc} just by keeping the linear in $\tau$ piece. We find for $\tau_0>0>\tau_0'$
\begin{align}
    &c_1 = \frac{2(\tau_0 + \tau_0' + \varphi_0 - \varphi_0')}{\pi},\quad \quad c_2 = \frac{\tau_0 - \tau_0'}2 + \frac{\varphi_0 + \varphi_0'}2- \log 2 \quad.
\end{align}
The corresponding $\varphi$ profile for these coefficients reaches an extrema at some point in between the initial and final slices, but such a point will not necessarily occur at $\tau = 0$ unless $a = a'$ ($\tau_0 = -\tau_0'$) and $\varphi_0 = \varphi_0'$ in which case $c_1 = 0$ and the solution in \eqref{eqn:varphigensol} becomes reflection symmetric about $\tau = 0$. In this case where $\tau_0 > 0 > \tau_0'$ or $\tau_0'>0>\tau_0$, the order $\sqrt{\epsilon}$ contribution to $\phi$ only modifies the on-shell action by a small amount. The on-shell action is then just 
\begin{align}
    iI_{\text{connected}} \supset V_{S^3} \int_{\sqrt{\frac{3}{V_*}} \tau_0'}^{\sqrt{\frac{3}{V_*}} \tau_0 } dt \left(3 a - 3 a \,\dot{a}^2 - V_* a^3\right) \approx \pm 4 \pi^2 \sqrt{\frac{V_*}3} \left( a_0^3 + a_0'^3\right) + \mathcal{O}(\epsilon) +O(a_0) \quad,
\end{align}
where we again expanded the first line of \eqref{eqn:D8}. The two signs correspond to the two different choices of orderings for $\tau_0$ and $\tau_0'$. These two saddles then just add within the gravitational path integral to give the ``cylinder" contribution to the path integral of the form 
\begin{align}
    \braket{a_0, \varphi_0}{a_0', \varphi_0'} \supset \ \sim\cos 4 \pi^2 \sqrt{\frac{V_*}3} \left( a_0^3 + a_0'^3\right).
\end{align}
where again we have ignored one-loop factors.

If the signs of $\tau_0$ and $\tau_0'$ are the same -- namely the two slices are on the same side of the bounce in the pure de Sitter solution -- then there are no solutions to the two equations in \eqref{eqn:phibc} unless we involve the higher order terms in $e^{-|\tau|}$ in \eqref{eqn:philargetau}. If we take the limit where $a_0, a_0'$ get large but $x = \frac{a_0'}{a_0}$ is held fixed then, assuming $\tau_0>\tau_0'>0$ so that $x<1$, we get
 \begin{small}
 \begin{equation}
\begin{aligned}
    c_1&\approx \frac{V_*^{3/2}(a_0')^3}{\sqrt{3}}\frac{\varphi_0-\varphi_0'-\log x}{1-x^3}+\frac{\sqrt{3V_*}a_0'}{4}\left[\frac{1+x}{1+x+x^2}-\frac{3(1+x+x^2+x^3+x^4)(\varphi_0-\varphi_0'-\log x)}{(1-x)(1+x+x^2)^2}\right]\\ &+O\left(\frac{1}{a_0'}\right)\\
    c_2&\approx -\frac{\pi V_*^{3/2}(a_0')^3}{4\sqrt{3}}\frac{\varphi_0-\varphi_0'-\log x}{1-x^3}-\frac{\sqrt{3V_*}\pi a_0'}{16}\left[\frac{4(1+x)}{1+x+x^2}-\frac{3(1+x+x^2+x^3+x^4)(\varphi_0-\varphi_0'-\log x)}{(1-x)(1+x+x^2)^2}\right]\\
    &\quad+\frac{\varphi_0-x^3\varphi_0'-\log x}{1-x^3}+\log\left(a_0'\sqrt{V_*/3}\right)+O\left(\frac{1}{(a_0')^{-2}}\right)
    \end{aligned}
\end{equation}
\end{small}
In this limit, $\varphi$ takes the form
\begin{align}\label{eqn:phisol}
    \varphi(\tau) = \log a_0'    - \log a(\tau) - \frac{a'^3_0 (\varphi_0 - \varphi_0') + a^3(\tau)(x^3 \varphi_0' - \varphi_0) + (a^3(\tau)- a'^3_0) \log x}{a^3(\tau) (1-x^3 )}\quad,
\end{align}
with $a(\tau) = \sqrt{\frac{3}{V_*}} \cosh \tau$.

For this solution, where both slices sit on the same side of the bounce, we expect to get a delta function out of the final answer. One signal of this delta function is to look at the solution for $\varphi(\tau_0) \neq \varphi(\tau_0')$. A pole in the action as $a_0 \to a_0'$  with $\varphi(\tau_0), \varphi(\tau_0')$ held fixed is a signal that the full path integral is outputting a delta function. To see that the full path integral diverges when $\varphi_0 = \varphi_0'$ and thus obtain a delta function, one should compute the one-loop determinant, which would yield a diverging pre-factor for $\varphi_0\to\varphi_0'$. Plugging in the solution in \eqref{eqn:phisol} to the matter action, expanding to linear order in $\epsilon$ and then taking the limit where $a_0, a_0'$ are large but $a_0'/a_0 = x$ is taken toward one, we get\footnote{Notice that there we are neglecting $1/a_0$ corrections at order $\epsilon^0$, because they are not divergent in the $a_0\to a_0'$ limit of interest and therefore subleading with respect to the $O(\epsilon)$ term we included.}
\begin{align}\label{eqn:disconnectedapp}
    &iI_{\text{connected}} \approx \pm \left(4 \pi^2 \sqrt{\frac{V_*}3} \left( a_0^3 - a_0'^3\right) + \epsilon \pi^2\sqrt{\frac{V_*}3} \int_{a_0'}^{a_0} da \ a^4 \left(\frac{d \varphi}{d a}\right)^2 \right) \nonumber \\
    & \approx\pm \left(4 \pi^2 \sqrt{\frac{V_*}3} \left( a_0^3 - a_0'^3\right)+ \epsilon \pi^2 \sqrt{\frac{V_*}3} \frac{a_0^3(\varphi_0 - \varphi_0')^2}{1-x}  \right)\quad.
\end{align}
The two signs correspond to the two slices being both after or before the bounce at $t = 0$. Note that this has a pole at $x = 1$ or $a_0 = a_0'$. This is a signal that the overlap between the states is becoming a delta function in the inflaton value, namely
\begin{align}
   \lim_{a_0 \to a_0'} \braket{a_0,\varphi_0}{a_0', \varphi_0'} \supset \sim \delta(\varphi_0 - \varphi_0').
\end{align}
Again, to see that the answer actually diverges when $\varphi_0 = \varphi_0'$, one would need to perform a one-loop analysis. 

\subsubsection*{On-shell Action for Conformal Boundary Conditions $\gamma=1/3$}

For the boundary at $t=t_0$ we need to add the following term to Eq. \eqref{eqn:D7}
 \begin{equation}
     \frac{2}{3}\int_{\partial M}K=2V_{S^3}\left(a^2(t)\dot{a}(t)\right)|_{t=t_0}\,.
     \label{eq:d18}
 \end{equation}
A similar term must be added for the boundary at $t=t_0'$.
Adding this to the first line of \eqref{eqn:D8} yields the on-shell action 
 \begin{align}\label{eqn:D19}
 iI &\supset \left. \frac{4\pi^2\sqrt{H_*^2 a^2(K)-1}}{H_*^2}\right\vert_{\partial M} = \left.\frac{4\pi^2\kappa}{H_*^2 \sqrt{1-\kappa^2}}\right\vert_{\partial M},
 \end{align}
where in the second equality we used $a(K) = \frac{1}{H_*\sqrt{1-\kappa^2}}$ with $K = 3H_* \kappa $ as in the main text. The inclusion symbol indicates that we should also include the term $\frac{\pm i 4\pi^2}{H_*^2}$ coming from no-boundary "boundaries" where $\kappa \to \pm i \infty$ when we compute disconnected contributions. Formula \eqref{eqn:D19} reproduces the order $\epsilon^0$ piece of \eqref{eqn:disconnected} and \eqref{eqn:osactionmain} in the main text. Notice that the action of the Euclidean portion of the geometry is clearly unaffected by the addition of \eqref{eq:d18}. Moreover, the order $\epsilon$ part of the action is also unaffected. The order $\epsilon$ part of Eq. \eqref{eqn:osactionmain} can then be obtained by replacing $a_0, a_0'$ by the corresponding functions of $K$ in the order $\epsilon$ piece of \eqref{eqn:disconnectedapp}.

\bibliographystyle{jhep}
\bibliography{references}
\end{document}